\documentclass[letterpaper,12pt]{article}
\usepackage[margin=2cm]{geometry}
\usepackage{amsmath}
\usepackage{amssymb}
\usepackage[affil-it]{authblk}
\usepackage[labelfont=bf]{caption}
\usepackage{float}
\usepackage{fancyhdr}
\usepackage{indentfirst}
\usepackage[backend=biber,style=numeric-comp,sorting=none,doi=true,url=false]{biblatex}
\usepackage[colorlinks=true,linkcolor=red,citecolor=blue,urlcolor=blue]{hyperref}
\allowdisplaybreaks
\title{A Time-Dependent Canonical Transformation between Bateman and Doubled Caldirola--Kanai Systems for a Homogeneous Massive Scalar Field on a Prescribed FLRW Background}

\author[1]{Narakorn Kaewkhao\thanks{E-mail: naragorn.k@psu.ac.th}}
\author[1]{Chaiyaphat Phantusen\thanks{E-mail: chaiyaphat.pts@gmail.com}}

\affil[1]{Division of Physical Science (Physics), Faculty of Science, Prince of Songkla University, Hat Yai, Songkhla 90110, Thailand}

\date{\today}

\begin{document}
\maketitle

\hrule
\begin{abstract}
Dissipative equations admit distinct variational descriptions in the Bateman and Caldirola--Kanai (CK) formalisms. The classical correspondence between them is extended to a homogeneous canonical scalar field on a prescribed spatially flat Friedmann--Lema\^itre--Robertson--Walker (FLRW) background, where the expansion produces the time-dependent damping coefficient $3H(t)$. A multiplier action yields the Klein--Gordon equation and a complementary anti-damped equation containing the term $-3\dot{H}(t)\chi$. A first-order Bateman Lagrangian derived from the same multiplier action reproduces this physical--auxiliary pair for a general potential. Specializing to a free massive field gives the Bateman and doubled CK Lagrangians and Hamiltonians used in the canonical comparison. The factors $a^{3}(t)$ and $a^{-3}(t)$ generate the damped and anti-damped CK sectors, respectively. An explicit time-dependent canonical transformation, generated by a function linear in the Bateman momenta, maps the complete doubled CK system to the Bateman system. For this point transformation, the terms proportional to $\dot{H}(t)$ are required for Hamiltonian equivalence. In rotated variables, the Bateman scalar-field Hamiltonian takes the difference form $H_{B,\mathrm{SF}} = E_{u} - E_{v}$. It is conserved for constant $H$ and generally varies with time otherwise. For the power-law background $a(t) \propto t^{p}$, however, a correlated family at $p = 2/3$ has conserved $H_{B,\mathrm{SF}}$ despite the time dependence of $H(t)$. These results concern classical homogeneous fields on a prescribed FLRW background and exclude the gravitational phase space.
\end{abstract}

\section{Introduction}
\label{sec:introduction}

Variational descriptions of dissipative dynamics often require extensions of the usual conservative-system framework~\cite{Bauer1931,Galley2013,Dodin2017}. For the damped harmonic oscillator, Bateman introduced an auxiliary variable satisfying the corresponding amplified equation and thereby obtained an autonomous Hamiltonian for the enlarged system~\cite{Bateman1931}. Caldirola and Kanai used a different approach in which an explicitly time-dependent factor in the Lagrangian produces the damped equation without an additional coordinate~\cite{Caldirola1941, Kanai1948}. Although the Bateman and Caldirola--Kanai (CK) formulations describe the same damped motion, they use different phase spaces and assign different roles to their Hamiltonians.

Several relations between these formulations have been established. Dekker showed that the time-independent Bateman Hamiltonian can be related to the time-dependent CK Hamiltonian after a restriction is imposed on the classical trajectories~\cite{Dekker1981}. A time-independent Hamiltonian description based on first integrals has also been constructed for a single damped oscillator, together with canonical transformations to standard Hamiltonian forms~\cite{Chandrasekar2007}. Cariglia et al.~\cite{Cariglia2016} constructed a canonical transformation between a doubled CK system and the Bateman system in the course of studying Eisenhart lifts and Arnold transformations. Their transformation acts on phase space and is not induced by a transformation of the configuration variables. Guerrero et al.~\cite{Guerrero2011} obtained the Bateman system by extending the symmetry algebra of the CK oscillator and recovered the CK description through time-dependent constraints. Schuch et al.~\cite{Schuch2015} related the Bateman system to expanding coordinates by eliminating the auxiliary variables through nonunique constraints, after which the expanding-coordinate formulation was connected to the CK system. These approaches show that the relation between the two descriptions depends on how the auxiliary sector and the canonical variables are introduced.

Time-dependent configuration-space dilatations have also been used as quantum canonical transformations for the CK oscillator~\cite{Mostafazadeh1998}. Here, the generating function is restricted to a linear time-dependent point transformation: it is linear in the Bateman momenta, so the transformed coordinates depend only on the CK coordinates and time. This form permits a direct comparison of the configuration-space Lagrangians through a boundary term. Replacing the exponential rescalings with powers of $a(t)$ extends the same structure to a prescribed FLRW background, with the $\dot{H}(t)$ terms retained as required by Hamiltonian equivalence. This restriction distinguishes the resulting map from the more general phase-space transformation of Cariglia et al.~\cite{Cariglia2016}.

A similar damping structure occurs for a homogeneous scalar field in a spatially flat Friedmann--Lema\^itre--Robertson--Walker (FLRW) spacetime. The Klein--Gordon equation contains the term $3H(t)\dot{\phi}$, which represents damping due to the expansion of the background geometry~\cite{Cariglia2016, Copeland2006}. Dissipative scalar-field models have also been considered in other settings. CK-type constructions have been applied to dissipative scalar fields~\cite{Jafari2021,AlvesJunior2025}, while bilinear and doubled-field actions have been used to formulate linearly damped scalar-field theories~\cite{Trachenko2019,Baggioli2020,SahaAashish2026}. Warm-inflation models, for example, obtain dissipation and radiation production from interactions between the scalar field and other degrees of freedom~\cite{BereraRamos2003}. Harko~\cite{Harko2023} introduced an exponential dissipation factor into the scalar-field Lagrangian and studied the resulting modifications of the Klein--Gordon and Friedmann equations. El-Nabulsi and Anukool~\cite{ElNabulsi2026} applied a Bateman-type dual-field structure to tachyonic and Born--Infeld or Dirac--Born--Infeld scalar models. Here, the physical field remains a minimally coupled canonical scalar field, and the scale factor is treated as a prescribed background function.

The question addressed in this work is whether the canonical correspondence between the Bateman and doubled CK systems can be extended consistently to a scalar field with the time-dependent damping coefficient $3H(t)$. The principal result is an explicit time-dependent canonical map, induced by a linear point transformation, between the complete doubled Bateman and CK scalar-field systems on a sufficiently differentiable prescribed FLRW background. For the linearized second-order operator with coefficients $r(t)$ and $s(t)$, the formal adjoint replaces them by $-r(t)$ and $s(t) - \dot{r}(t)$, respectively~\cite{Talukdar2020}. In the scalar-field problem, $r(t) = 3H(t)$ and $s(t) = V''(\phi(t))$, so the auxiliary equation contains the term $-3\dot{H}(t)\chi$. The multiplier construction is first developed for a general potential, after which the free massive potential is selected to obtain the linear doubled system used in the canonical analysis.

The classical damped oscillator is first used to establish the required canonical structure. In the rotated variables, the Bateman Hamiltonian takes the form $H_{B} = E_{u} - E_{v}$, which displays its indefinite character without identifying it with the ordinary mechanical energies of the original damped and amplified coordinates. A doubled CK system is then constructed, and the corresponding generating function and canonical map are derived explicitly.

For the cosmological system, the multiplier action gives the physical Klein--Gordon equation and a complementary anti-damped equation for an auxiliary field. A first-order Bateman Lagrangian derived from the same multiplier action reproduces both equations for a general potential. The free massive potential is then selected for the explicit Hamiltonian construction. The resulting Bateman Hamiltonian is conserved for constant $H$ but not generically when $H(t)$ varies. For a power-law background $a(t) \propto t^{p}$, the compatibility condition admits a nontrivial correlated family at $p = 2/3$, providing an explicit case in which $H(t)$ is time dependent while $H_{B,\mathrm{SF}}$ is conserved. The damped and anti-damped equations are also derived from CK Lagrangians with the factors $a^{3}(t)$ and $a^{-3}(t)$, respectively. An explicit time-dependent canonical transformation then relates the complete doubled CK and Bateman systems, including the terms generated by $\dot{H}(t)$. The construction applies to homogeneous classical fields on a prescribed FLRW background and does not include the gravitational phase space or modify the Friedmann equations.
\section{Bateman Dual-System and Caldirola--Kanai Formulations}
\label{sec:classical_bateman_ck_formulations}

The classical damped harmonic oscillator is described first in the Bateman dual-system and doubled Caldirola--Kanai formulations. A time-dependent canonical transformation between the two Hamiltonian descriptions is then extended to the cosmological scalar-field system in the following sections.

\subsection{Bateman's Dual-System Approach}
\label{subsec:bateman_dual_system}
In 1931, H. Bateman~\cite{Bateman1931} formulated a variational description of dissipative systems by introducing a complementary equation of motion. The equation of motion for $x(t)$ can be written as
\begin{equation}
    \ddot{x} + \gamma \dot{x} + \omega^{2} x = 0, \label{EOM_x}
\end{equation}
where $\gamma \geq 0$ is a constant damping coefficient and $\omega > 0$ is the natural angular frequency of the undamped oscillator. The coefficient of $\dot{x}$ is denoted by $\gamma$, rather than Bateman's original notation $2k$, following the convention commonly used in later studies~\cite{Cariglia2016, Guerrero2011, Bagarello2019, Vestal2021, Deguchi2020, Deguchi2019, Blasone2004, Schuch2015}. Other conventions include a damping term written as $2\gamma\dot{x}$~\cite{Chruscinski2006} or $2\lambda\dot{x}$~\cite{JavierValdez2025}.

The variational construction begins by introducing a function $y(t)$ as a Lagrange multiplier for Eq.~\eqref{EOM_x}. The corresponding multiplier action is
\begin{equation}
    S_{M}[x,y] \equiv m \int_{t_{i}}^{t_{f}} y \left(\ddot{x} + \gamma\dot{x} + \omega^{2}x\right)dt. \label{action_Bateman_multiplier}
\end{equation}
Here, $m > 0$ denotes the oscillator mass. Its inclusion fixes the multiplier normalization so that $y$ has the same physical dimension as $x$. The quantities $t_{i}$ and $t_{f}$ denote the initial and final times, respectively. Variation of Eq.~\eqref{action_Bateman_multiplier} with respect to $y$ gives the damped equation \eqref{EOM_x}. Integrating the term $my\ddot{x}$ by parts gives
\begin{equation}
    S_{M}[x,y] = \left[my\dot{x}\right]_{t_{i}}^{t_{f}} + S_{1}[x,y], \qquad S_{1}[x,y] \equiv \int_{t_{i}}^{t_{f}} L_{1}\,dt, \label{relation_multiplier_first_order}
\end{equation}
where the first-order multiplier Lagrangian is
\begin{equation}
    L_{1} = -m\dot{x}\dot{y} + m\gamma y\dot{x} + m\omega^{2}xy. \label{lagrangian_Bateman_multiplier}
\end{equation}
The Euler--Lagrange equation obtained by varying $L_{1}$ with respect to $y$ reproduces Eq.~\eqref{EOM_x}, while variation with respect to $x$ gives
\begin{equation}
    \ddot{y} - \gamma\dot{y} + \omega^{2}y = 0. \label{EOM_y}
\end{equation}
Although $y(t)$ is introduced as a Lagrange multiplier, the first-order action treats $x(t)$ and $y(t)$ as independent variables, and variation with respect to $x$ determines the evolution of $y$. This variable has been described as a real coordinate~\cite{Deguchi2019}, a position-like or dual variable~\cite{Schuch2015}, an auxiliary field~\cite{Cariglia2016}, an auxiliary variable~\cite{JavierValdez2025}, or simply a coordinate variable~\cite{Vestal2021}. The term auxiliary variable is adopted below.

The differential operator in Eq.~\eqref{EOM_y} is the formal adjoint of that in Eq.~\eqref{EOM_x}. The pair therefore provides an indirect variational representation of the damped oscillator and its complementary degree of freedom~\cite{Talukdar2020}. Equation~\eqref{EOM_y} describes an amplified, or equivalently anti-damped, oscillator~\cite{Cariglia2016, Bagarello2019, Vestal2021, Deguchi2020, Pal2018, Deguchi2019, Chruscinski2006, Blasone2004}. It may also be regarded as the time-reversed counterpart of the damped oscillator~\cite{Guerrero2011, JavierValdez2025, Vestal2021, Pal2018, Blasone2004, Schuch2015}. In some physical interpretations, the $y$-degree of freedom represents an effective reservoir or the absorbing member of the complementary pair~\cite{Guerrero2011, Blasone2004}. No reservoir interpretation is assigned to $y$ here; it is treated only as an auxiliary variable governed by Eq.~\eqref{EOM_y}.

The system described by Eq.~\eqref{EOM_y} will be called the amplified harmonic oscillator. It differs from Eq.~\eqref{EOM_x} only in the sign of the damping term. For $\gamma > 0$, its mechanical energy increases; in the underdamped regime $\omega > \gamma/2$, the oscillatory solutions have an envelope proportional to $e^{\gamma t/2}$~\cite{Deguchi2019}.

The boundary term is chosen so that the damping-dependent velocity coupling takes the standard Bateman form. A first-order Bateman Lagrangian is then defined by
\begin{equation}
    L_{B} \equiv -L_{1} + \frac{d}{dt}\left(\frac{m\gamma}{2}xy\right). \label{relation_Lagrangians}
\end{equation}
Evaluation of the total derivative gives
\begin{equation}
    L_{B} = m\dot{x}\dot{y} + \frac{m\gamma}{2}\left(x\dot{y}-y\dot{x}\right) - m\omega^{2}xy. \label{lagrangian_Bateman}
\end{equation}
The corresponding Bateman action is
\begin{equation}
    S_{B}[x,y] \equiv \int_{t_{i}}^{t_{f}} L_{B}\,dt. \label{action_Bateman}
\end{equation}
At the action level, Eqs.~\eqref{relation_multiplier_first_order} and \eqref{relation_Lagrangians} give
\begin{equation}
    S_{B}[x,y] = -S_{1}[x,y] + \left[\frac{m\gamma}{2}xy\right]_{t_{i}}^{t_{f}} = -S_{M}[x,y] + \left[my\dot{x} + \frac{m\gamma}{2}xy\right]_{t_{i}}^{t_{f}}. \label{relation_actions_Bateman}
\end{equation}
The first equality in Eq.~\eqref{relation_actions_Bateman} shows that $S_{B}$ and $S_{1}$ differ by an overall sign and a boundary term depending only on the endpoint coordinates. They therefore yield the same Euler--Lagrange equations under fixed-endpoint variations. Because $S_{M}$ contains $\ddot{x}$, its direct variation with respect to $x$ may be formulated by imposing $\delta x = \delta \dot{x} = 0$ at $t_{i}$ and $t_{f}$. The term $my\dot{x}$ appears in the boundary contribution because the multiplier action contains $\ddot{x}$. The first-order action $S_{1}$ is consequently the appropriate variational representative for constructing the canonical formulation. The conjugate momenta obtained from $L_{B}$ are
\begin{align}
    p_{x} \equiv \frac{\partial L_{B}}{\partial \dot{x}} = m \dot{y} - \frac{\gamma m y}{2}
    \quad &\Rightarrow \quad
    \dot{y} = \frac{p_{x}}{m} + \frac{\gamma y}{2}, \label{momentum_Bateman_x}
    \\
    p_{y} \equiv \frac{\partial L_{B}}{\partial \dot{y}} = m \dot{x} + \frac{\gamma m x}{2}
    \quad &\Rightarrow \quad
    \dot{x} = \frac{p_{y}}{m} - \frac{\gamma x}{2}. \label{momentum_Bateman_y}
\end{align}
The Legendre transform of $L_{B}$ gives the Bateman Hamiltonian,
\begin{align}
    H_{B} &= \dot{x} p_{x} + \dot{y} p_{y} - L_{B}, \nonumber
    \\
    &= m \dot{x} \dot{y} + m \omega^{2} x y, \nonumber
    \\
    &= \frac{p_{x} p_{y}}{m} + \frac{\gamma}{2} \big(y p_{y} - x p_{x}\big) + m \Omega^{2} x y, \label{hamiltonian_Bateman}
\end{align}
where $\Omega^{2} \equiv \omega^{2} - \frac{\gamma^{2}}{4}$. In the underdamped regime, $\Omega^{2} > 0$, and $\Omega$ is the damped angular frequency. The conserved Bateman Hamiltonian should not be identified with the sum of the ordinary mechanical energies of the $x$- and $y$-oscillators. As discussed by Schuch~\cite{Schuch2015}, a direct energy-transfer interpretation would require relations between the two trajectories, such as $\dot{y} = \pm \dot{x}$, that do not hold for general solutions. Moreover, $H_{B}$ does not admit a natural decomposition of the conventional form $H_{B} = H_{x} + H_{y} + H_{xy}$ in the $(x,y)$ representation~\cite{Schuch2015}. The rotated variables introduced below make its indefinite structure more transparent.

A useful reformulation of the Bateman system is discussed by Cariglia et al.~\cite{Cariglia2016}. Although the equations of motion for $x$ and $y$ are dynamically decoupled, the same dynamics can be expressed as a pair of coupled oscillator equations by introducing linear combinations of these variables. Following the normalized coordinate convention used in Refs.~\cite{JavierValdez2025, Blasone2004}, the rotated variables $u$ and $v$ are defined by
\begin{align}
    u &= \frac{x + y}{\sqrt{2}}, \label{variable_u} 
    \\
    v &= \frac{x - y}{\sqrt{2}}. \label{variable_v}
\end{align}

Adding Eqs.~\eqref{EOM_x} and \eqref{EOM_y}, and rewriting the result in terms of the variables $u$ and $v$, gives the first equation of the coupled system:
\begin{align}
    \ddot{u} + \omega^{2} u = - \gamma \dot{v}. \label{EOM_u}
\end{align}
Multiplying Eq.~\eqref{EOM_u} by $m \dot{u}$ gives the time rate of change of the energy $E_{u}$:
\begin{align}
    \frac{dE_{u}}{dt} \equiv \frac{d}{dt} \left( \frac{1}{2} m \dot{u}^{2} + \frac{1}{2} m \omega^{2} u^{2} \right) = -m \gamma \dot{u} \dot{v}. \label{dE_u}
\end{align}

Similarly, subtracting Eq.~\eqref{EOM_y} from Eq.~\eqref{EOM_x}, and rewriting the result in terms of $u$ and $v$, gives the second equation of the coupled system:
\begin{align}
    \ddot{v} + \omega^{2} v = -\gamma \dot{u}. \label{EOM_v}
\end{align}
Multiplying Eq.~\eqref{EOM_v} by $m \dot{v}$ gives the time rate of change of the energy $E_{v}$:
\begin{align}
    \frac{d E_{v}}{dt} \equiv \frac{d}{dt} \left( \frac{1}{2} m \dot{v}^{2} + \frac{1}{2} m \omega^{2} v^{2} \right) = -m \gamma \dot{u} \dot{v}. \label{dE_v}
\end{align}

Equations~\eqref{dE_u} and \eqref{dE_v} show that the two mode-energy functions change at the same rate. In terms of the variables $u$ and $v$, the Bateman Hamiltonian can be written as $H_{B} = E_{u} - E_{v}$. It therefore follows that
\begin{align}
    \frac{d H_{B}}{dt} \equiv \frac{d E_{u}}{dt} - \frac{d E_{v}}{dt} = 0. 
    \label{dH_B}
\end{align}

Equation~\eqref{dH_B} shows that $H_{B}$ is conserved even though neither mode-energy function is conserved separately. Although both $E_{u}$ and $E_{v}$ are positive definite, their difference is indefinite and is not bounded from below at the classical level~\cite{Pal2018}. The same conservation law can be verified directly by evaluating the time derivative of $H_{B}$.

\subsection{Doubled Caldirola--Kanai Formulation of the Damped and Amplified Oscillators}
\label{subsec:doubled_ck_oscillators}
The Caldirola--Kanai formulation describes the damped harmonic oscillator by means of a single configuration variable and an explicitly time-dependent Lagrangian~\cite{Caldirola1941, Kanai1948}. Multiplying the ordinary harmonic-oscillator Lagrangian by $e^{\gamma t}$ yields the damped equation of motion without introducing the auxiliary coordinate required in the Bateman formulation.

The CK coordinates and their conjugate momenta are denoted by $(x', y')$ and $(p_{x'}, p_{y'})$, respectively, to distinguish them from the Bateman variables $(x, y, p_{x}, p_{y})$ introduced in the preceding subsection. Primes label the CK description and do not denote differentiation. A doubled CK system consisting of damped and amplified sectors is introduced for comparison with the Bateman formulation~\cite{Cariglia2016}.

For the damped coordinate $x'$, Eq.~\eqref{EOM_x} follows from the CK Lagrangian~\cite{Cariglia2016, Vestal2021}
\begin{align}
    L_{x'}(x', \dot{x}', t) &= e^{\gamma t} \left[ \frac{1}{2} m \left(\dot{x}'\right)^2 - \frac{1}{2} m \omega^{2} \left(x'\right)^2 \right], \nonumber
    \\
    &= \frac{1}{2} m e^{\gamma t} \left(\dot{x}'\right)^2 - \frac{1}{2} m \omega^{2} e^{\gamma t} \left(x'\right)^2. \label{lagrangian_CK_xprime}
\end{align}
Applying the Euler--Lagrange equation to $L_{x'}$ gives
\begin{equation*}
    \frac{d}{dt} \left( \frac{\partial L_{x'}}{\partial \dot{x}'} \right) - \frac{\partial L_{x'}}{\partial x'} = m e^{\gamma t} \left( \ddot{x}' + \gamma \dot{x}' + \omega^{2} x' \right) = 0.
\end{equation*}
Since $e^{\gamma t} \neq 0$, the Euler--Lagrange equation is Eq.~\eqref{EOM_x} with $x$ replaced by $x'$.

For the amplified coordinate $y'$, consider a CK-type Lagrangian that reproduces Eq.~\eqref{EOM_y}. A time-dependent multiplier $f_{y'}(t)$ is introduced in the form
\begin{equation*}
    L_{y'} = f_{y'}(t) \left[ \frac{1}{2} m \left(\dot{y}'\right)^2 - \frac{1}{2} m \omega^{2} \left(y'\right)^2 \right].
\end{equation*}
The corresponding Euler--Lagrange equation is
\begin{equation*}
    \ddot{y}' + \frac{\dot{f}_{y'}}{f_{y'}} \dot{y}' + \omega^{2} y' = 0.
\end{equation*}
Matching Eq.~\eqref{EOM_y} requires $\dot{f}_{y'}/f_{y'} = -\gamma$, and hence $f_{y'}(t) = Ce^{-\gamma t}$. The constant $C \neq 0$ does not affect the Euler--Lagrange equation, and the normalization $C = 1$ is adopted. The amplified-sector CK Lagrangian is then
\begin{align}
    L_{y'}(y', \dot{y}', t) &= e^{-\gamma t} \left[ \frac{1}{2} m \left(\dot{y}'\right)^2 - \frac{1}{2} m \omega^{2} \left(y'\right)^2 \right], \nonumber
    \\
    &= \frac{1}{2} m e^{-\gamma t} \left(\dot{y}'\right)^2 - \frac{1}{2} m \omega^{2} e^{-\gamma t} \left(y'\right)^2. \label{lagrangian_CK_yprime}
\end{align}

The damped and amplified sectors are combined with a relative minus sign. This choice leaves their Euler--Lagrange equations unchanged and gives the doubled CK system the indefinite difference structure used in its canonical relation to the Bateman Hamiltonian~\cite{Cariglia2016}:
\begin{equation}
    L_{CK} = L_{x'} - L_{y'} = \frac{1}{2} m \left[ e^{\gamma t} \left(\dot{x}'\right)^2 - e^{-\gamma t} \left(\dot{y}'\right)^2 \right]
    - \frac{1}{2} m \omega^{2} \left[ e^{\gamma t} \left(x'\right)^2 - e^{-\gamma t} \left(y'\right)^2 \right]. \label{lagrangian_CK_total}
\end{equation}
From this doubled Lagrangian, the canonical momenta conjugate to $x'$ and $y'$ are defined by
\begin{alignat}{2}
    p_{x'} &\equiv \frac{\partial L_{CK}}{\partial \dot{x}'} = m e^{\gamma t} \dot{x}'
    \quad &\Rightarrow \quad
    \dot{x}' &= \frac{p_{x'}}{m} e^{-\gamma t}, \label{momentum_CK_xprime}
    \\
    p_{y'} &\equiv \frac{\partial L_{CK}}{\partial \dot{y}'} = -m e^{-\gamma t} \dot{y}'
    \quad &\Rightarrow \quad
    \dot{y}' &= -\frac{p_{y'}}{m} e^{\gamma t}. \label{momentum_CK_yprime}
\end{alignat}
The Legendre transform of $L_{CK}$ gives
\begin{align}
    H_{CK} &= \dot{x}' p_{x'} + \dot{y}' p_{y'} - L_{CK}, \nonumber 
    \\
    &= \frac{e^{-\gamma t}}{2m} p_{x'}^2 - \frac{e^{\gamma t}}{2m} p_{y'}^2 + \frac{1}{2} m \omega^{2} \left[ e^{\gamma t} \left(x'\right)^2 - e^{-\gamma t} \left(y'\right)^2 \right]. \label{hamiltonian_CK_total}
\end{align}
The Hamiltonian in Eq.~\eqref{hamiltonian_CK_total} can therefore be written as $H_{CK} = H_{x'} - H_{y'}$, where $H_{x'}$ and $H_{y'}$ denote the single-sector CK Hamiltonians for the damped and amplified oscillators, respectively.

The standard single-variable CK Hamiltonian for the damped oscillator is recovered by restricting the doubled system to the invariant sector $y' = p_{y'} = 0$. Equation~\eqref{hamiltonian_CK_total} then reduces to the standard CK Hamiltonian given in Eq.~(III.88) of Ref.~\cite{Cariglia2016}, Eq.~(20) of Ref.~\cite{Schuch2015}, and Kanai's original formulation~\cite{Kanai1948}:
\begin{equation}
    H_{CK}^{(x')}(t, x', p_{x'}) = \frac{e^{-\gamma t}}{2m} p_{x'}^2 + \frac{1}{2} m \omega^{2} e^{\gamma t} \left(x'\right)^2. 
    \label{hamiltonian_CK_xprime}
\end{equation}
\subsection{Canonical Transformation between the Doubled CK and Bateman Systems}
\label{subsec:classical_ck_bateman_transformation}

The difference form $H_{CK} = H_{x'} - H_{y'}$ agrees with Eq.~(III.117) of Cariglia et al.~\cite{Cariglia2016}. A canonical transformation induced by a time-dependent point transformation is constructed by expressing the Bateman variables in terms of the CK variables and requiring the two configuration-space Lagrangians to differ by a total time derivative:
\begin{equation}
    L_{CK} = L_{B} + \frac{dF_{1}}{dt}, 
    \label{relation_Lagrangians_Bateman_CK}
\end{equation}
where $F_{1}$ is the boundary function associated with the transformation. Using the Legendre-transform identities for the doubled CK and Bateman systems, Eq.~\eqref{relation_Lagrangians_Bateman_CK} takes the first-order form
\begin{equation}
    p_{x'}\dot{x}' + p_{y'}\dot{y}' - H_{CK} = p_x\dot{x} + p_y\dot{y} - H_B + \frac{dF_1}{dt}.
    \label{relation_Hamiltonian_Bateman_CK}
\end{equation}
A type-2 generating function is introduced as
\begin{equation}
    F_{2}(x', y', p_{x}, p_{y},t) = F_{1} + x p_{x} + y p_{y}.
    \label{F2def}
\end{equation}
Using $F_{1} = F_{2} - x p_{x} - y p_{y}$ in Eq.~\eqref{relation_Hamiltonian_Bateman_CK} and expanding its total time derivative gives
\begin{align}
    \left(p_{x'} - \frac{\partial F_{2}}{\partial x'}\right) \dot{x}'
    + \left(p_{y'} - \frac{\partial F_{2}}{\partial y'}\right) \dot{y}'
    + \left(x - \frac{\partial F_{2}}{\partial p_{x}}\right) \dot{p}_x
    + \left(y - \frac{\partial F_{2}}{\partial p_{y}}\right) \dot{p}_y
    = H_{CK} - H_{B} + \frac{\partial F_{2}}{\partial t}.
    \label{relation_Hamiltonian_Bateman_CK_expanded}
\end{align}
Since Eq.~\eqref{relation_Hamiltonian_Bateman_CK_expanded} must hold as an identity in the independent quantities $\dot{x}'$, $\dot{y}'$, $\dot{p}_{x}$, and $\dot{p}_{y}$, their coefficients must vanish separately. The resulting canonical-transformation conditions are
\begin{equation}
\begin{aligned}
    p_{x'} &= \frac{\partial F_{2}}{\partial x'},
    &\qquad
    p_{y'} &= \frac{\partial F_{2}}{\partial y'},
    \\
    x &= \frac{\partial F_{2}}{\partial p_{x}},
    &\qquad
    y &= \frac{\partial F_{2}}{\partial p_{y}}.
\end{aligned}
    \label{cano_trans_conditions}
\end{equation}
The remaining Hamiltonian condition is
\begin{equation}
    H_{B} = H_{CK} + \frac{\partial F_{2}}{\partial t}.
    \label{relation_Hamiltonians_Bateman_CK}
\end{equation}
Equation~\eqref{relation_Hamiltonians_Bateman_CK} coincides with the canonical relation used in Eq.~(III.118) of Ref.~\cite{Cariglia2016}. The generating function given in Eq.~(III.119) of Ref.~\cite{Cariglia2016} contains a term quadratic in the Bateman momenta. The resulting transformation therefore acts on phase space and does not follow from a point transformation of the configuration variables. The present construction instead uses a time-dependent point transformation in which the Bateman coordinates depend only on the CK coordinates and time:
\begin{equation*}
    x = x(x', y', t),
    \qquad
    y = y(x', y', t).
\end{equation*}
The corresponding type-2 generating function is therefore linear in the Bateman momenta:
\begin{equation}
    F_{2}(x', y', p_{x}, p_{y}, t) = A(x', y', t)\,p_{x} + B(x', y', t)\,p_{y} + G(x', y', t).
    \label{F2_ansatz}
\end{equation}
Time-dependent linear canonical transformations provide a general framework for Hamiltonians of harmonic-oscillator type~\cite{Leach1977}. The homogeneous quadratic form of both Hamiltonians motivates a linear phase-space transformation that leaves the origin fixed. Under this restriction, $A$ and $B$ are linear in $x'$ and $y'$, while $G$ is homogeneous quadratic:
\begin{align}
    A &= \alpha_{1}(t) x' + \beta_{1}(t) y',
    \qquad
    B = \alpha_{2}(t) x' + \beta_{2}(t) y',
    \label{eq_AB}
    \\
    G &= \frac{1}{2} g_{11}(t) (x')^2 + g_{12}(t) x' y' + \frac{1}{2} g_{22}(t) (y')^2.
    \label{eq_G}
\end{align}
Applying the canonical conditions in Eq.~\eqref{cano_trans_conditions} gives
\begin{align}
    p_{x'} &= \alpha_{1} p_{x} + \alpha_{2} p_{y} + g_{11} x' + g_{12} y',
    \label{pxprime_transform}
    \\
    p_{y'} &= \beta_{1} p_{x} + \beta_{2} p_{y} + g_{12} x' + g_{22} y',
    \label{pyprime_transform}
    \\
    x &= \alpha_{1} x' + \beta_{1} y',
    \label{x_transform}
    \\
    y &= \alpha_{2} x' + \beta_{2} y'.
    \label{y_transform}
\end{align}
The explicit time derivative of the generating function is
\begin{align*}
    \frac{\partial F_{2}}{\partial t} &= \dot{\alpha}_{1} x' p_{x} + \dot{\beta}_{1} y' p_{x} + \dot{\alpha}_{2} x' p_{y} + \dot{\beta}_{2} y' p_{y} + \frac{1}{2} \dot{g}_{11} (x')^2 + \dot{g}_{12} x' y' + \frac{1}{2} \dot{g}_{22} (y')^2.
\end{align*}
Substituting Eqs.~\eqref{pxprime_transform}--\eqref{y_transform} and the preceding time derivative into Eq.~\eqref{relation_Hamiltonians_Bateman_CK}, and then collecting the ten independent monomials
\begin{equation*}
    p_{x}^{2}, \quad
    p_{y}^{2}, \quad
    p_{x}p_{y}, \quad
    (x')^{2}, \quad
    (y')^{2}, \quad
    x'y', \quad
    p_{x}x', \quad
    p_{x}y', \quad
    p_{y}x', \quad
    p_{y}y'.
\end{equation*}
gives the following system:
\begin{align}
    \frac{e^{-\gamma t}}{2m} \alpha_{1}^{2} - \frac{e^{\gamma t}}{2m} \beta_{1}^{2} &= 0,
    \label{coefficient_px2}
    \\
    \frac{e^{-\gamma t}}{2m} \alpha_{2}^{2} - \frac{e^{\gamma t}}{2m} \beta_{2}^{2} &= 0,
    \label{coefficient_py2}
    \\
    \frac{e^{-\gamma t}}{m} \alpha_{1}\alpha_{2} - \frac{e^{\gamma t}}{m} \beta_{1} \beta_{2} &= \frac{1}{m},
    \label{coefficient_pxpy}
    \\
    \frac{e^{-\gamma t}}{2m} g_{11}^{2} - \frac{e^{\gamma t}}{2m} g_{12}^{2} + \frac{m\omega^{2}}{2} e^{\gamma t} - m \Omega^{2} \alpha_{1} \alpha_{2} + \frac{1}{2} \dot{g}_{11} &= 0,
    \label{coefficient_xprime2}
    \\
    \frac{e^{-\gamma t}}{2m} g_{12}^{2} - \frac{e^{\gamma t}}{2m} g_{22}^{2} - \frac{m\omega^{2}}{2} e^{-\gamma t} - m \Omega^{2} \beta_{1} \beta_{2} + \frac{1}{2} \dot{g}_{22} &= 0,
    \label{coefficient_yprime2}
    \\
    \frac{e^{-\gamma t}}{m} g_{11}g_{12} - \frac{e^{\gamma t}}{m} g_{12} g_{22} - m \Omega^{2} \left(\alpha_{1} \beta_{2} + \alpha_{2} \beta_{1}\right) + \dot{g}_{12} &= 0,
    \label{coefficient_xprime_yprime}
    \\
    \frac{e^{-\gamma t}}{m} \alpha_{1} g_{11} - \frac{e^{\gamma t}}{m} \beta_{1} g_{12} + \dot{\alpha}_{1} + \frac{\gamma}{2} \alpha_{1} &= 0,
    \label{coefficient_px_xprime}
    \\
    \frac{e^{-\gamma t}}{m} \alpha_{1} g_{12} - \frac{e^{\gamma t}}{m} \beta_{1} g_{22} + \dot{\beta}_{1} + \frac{\gamma}{2} \beta_{1} &= 0,
    \label{coefficient_px_yprime}
    \\
    \frac{e^{-\gamma t}}{m} \alpha_{2} g_{11} - \frac{e^{\gamma t}}{m} \beta_{2} g_{12} + \dot{\alpha}_{2} - \frac{\gamma}{2} \alpha_{2} &= 0,
    \label{coefficient_py_xprime}
    \\
    \frac{e^{-\gamma t}}{m} \alpha_{2} g_{12} - \frac{e^{\gamma t}}{m} \beta_{2} g_{22} + \dot{\beta}_{2} - \frac{\gamma}{2} \beta_{2} &= 0.
    \label{coefficient_py_yprime}
\end{align}
Equations~\eqref{coefficient_px2} and \eqref{coefficient_py2} imply
\begin{equation*}
    \beta_{i} = s_{i} e^{-\gamma t} \alpha_{i},
    \qquad s_{i} \in \{+1, -1\},
    \qquad i = 1, 2.
\end{equation*}
The branches with $s_{1} = s_{2}$ eliminate the required $p_{x}p_{y}$ term and are excluded. The two remaining branches are related by the simultaneous sign reversal $(y', p_{y'}) \rightarrow (-y', -p_{y'})$. Choosing $s_{1} = +1$ and $s_{2} = -1$ gives
\begin{equation}
    \beta_{1} = e^{-\gamma t} \alpha_{1},
    \qquad
    \beta_{2} = -e^{-\gamma t}\alpha_{2}.
    \label{beta_1_2}
\end{equation}
Substitution into Eq.~\eqref{coefficient_pxpy} gives
\begin{equation}
    \alpha_{1}(t) \alpha_{2}(t) = \frac{1}{2} e^{\gamma t}.
    \label{alpha_1_2}
\end{equation}
Using Eq.~\eqref{beta_1_2} in Eqs.~\eqref{coefficient_px_xprime}--\eqref{coefficient_py_yprime} gives
\begin{equation}
    g_{12}(t) = 0,
    \label{g_12}
\end{equation}
together with
\begin{equation*}
    \frac{\dot{\alpha}_{2}}{\alpha_{2}} - \frac{\dot{\alpha}_{1}}{\alpha_{1}} = \gamma.
\end{equation*}
A second relation follows by differentiating Eq.~\eqref{alpha_1_2} with respect to time:
\begin{equation*}
    \frac{\dot{\alpha}_{1}}{\alpha_{1}} + \frac{\dot{\alpha}_{2}}{\alpha_{2}} = \gamma.
\end{equation*}
Combining these relations gives
\begin{equation}
    \frac{\dot{\alpha}_{1}}{\alpha_{1}} = 0,
    \qquad
    \frac{\dot{\alpha}_{2}}{\alpha_{2}} = \gamma.
    \label{alpha_dot}
\end{equation}
Hence, $\alpha_{1} = c \neq 0$ and Eq.~\eqref{alpha_1_2} gives $\alpha_{2} = e^{\gamma t}/(2c)$. Setting $c = 1/\sqrt{2}$ fixes the remaining normalization freedom and recovers the normalized sum-and-difference transformation in the limit $\gamma \to 0$:
\begin{equation}
    \alpha_{1} = \frac{1}{\sqrt{2}},
    \qquad
    \alpha_{2} = \frac{1}{\sqrt{2}} e^{\gamma t}.
\end{equation}
Equation~\eqref{beta_1_2} then gives
\begin{equation}
    \beta_{1} = \frac{1}{\sqrt{2}} e^{-\gamma t},
    \qquad
    \beta_{2} = -\frac{1}{\sqrt{2}}.
    \label{betas}
\end{equation}
The mixed-coefficient equations \eqref{coefficient_px_xprime}--\eqref{coefficient_py_yprime} also determine the diagonal coefficients:
\begin{equation}
    g_{11} = -\frac{m\gamma}{2} e^{\gamma t},
    \qquad
    g_{22} = -\frac{m\gamma}{2} e^{-\gamma t}.
    \label{gdiag}
\end{equation}
Finally, Eqs.~\eqref{coefficient_xprime2}--\eqref{coefficient_xprime_yprime} are satisfied identically upon using $\Omega^{2} = \omega^{2} - \gamma^{2}/4$. The resulting generating function is
\begin{align}
    F_{2} &= \frac{1}{\sqrt{2}} \left(x' + e^{-\gamma t} y'\right) p_{x} + \frac{1}{\sqrt{2}} \left(e^{\gamma t} x' - y'\right) p_{y} - \frac{m\gamma}{4} \left(e^{\gamma t} (x')^2 + e^{-\gamma t} (y')^2 \right).
    \label{F2P_final}
\end{align}
The corresponding transformation from the CK phase-space variables to the Bateman phase-space variables is
\begin{align}
    x &= \frac{x' + e^{-\gamma t} y'}{\sqrt{2}},
    \label{coordinate_transformation_x}
    \\
    y &= \frac{e^{\gamma t} x' - y'}{\sqrt{2}},
    \label{coordinate_transformation_y}
    \\
    p_{x} &= \frac{p_{x'} + e^{\gamma t} p_{y'}}{\sqrt{2}} + \frac{m\gamma}{2\sqrt{2}} \left(e^{\gamma t}x' + y'\right),
    \label{momentum_transformation_x}
    \\
    p_{y} &= \frac{e^{-\gamma t} p_{x'} - p_{y'}}{\sqrt{2}} + \frac{m\gamma}{2\sqrt{2}} \left(x' - e^{-\gamma t}y'\right).
    \label{momentum_transformation_y}
\end{align}
The inverse transformation from the Bateman phase-space variables to the CK phase-space variables is
\begin{align}
    x' &= \frac{x + e^{-\gamma t} y}{\sqrt{2}},
    \label{coordinate_transformation_xprime}
    \\
    y' &= \frac{e^{\gamma t} x - y}{\sqrt{2}},
    \label{coordinate_transformation_yprime}
    \\
    p_{x'} &= \frac{p_{x} + e^{\gamma t} p_{y}}{\sqrt{2}} - \frac{m\gamma}{2\sqrt{2}} \left(e^{\gamma t}x + y\right),
    \label{momentum_transformation_xprime}
    \\
    p_{y'} &= \frac{e^{-\gamma t} p_{x} - p_{y}}{\sqrt{2}} - \frac{m\gamma}{2\sqrt{2}} \left(x - e^{-\gamma t}y\right).
    \label{momentum_transformation_yprime}
\end{align}

The canonical character of the transformation can be checked directly from the equal-time Poisson brackets. At fixed $t$, the factors $e^{\pm\gamma t}$ are treated as coefficients. The symbols $\{\cdot, \cdot\}_{B}$ and $\{\cdot, \cdot\}_{CK}$ denote Poisson brackets evaluated with respect to the Bateman variables $(x, y, p_{x}, p_{y})$ and the CK variables $(x', y', p_{x'}, p_{y'})$, respectively. Using the inverse transformation for the primed variables and the direct transformation for the unprimed variables, the six independent brackets in each phase-space description are summarized in Table~\ref{table:canonical_Poisson_brackets}.
\begin{table}[H]
    \centering
    \renewcommand{\arraystretch}{1.35}
    \begin{tabular}{c c}
        \hline
        CK variables & Bateman variables
        \\
        \hline
        $\{x', y'\}_{B} = 0$ & $\{x, y\}_{CK} = 0$
        \\
        $\{x', p_{x'}\}_{B} = 1$ & $\{x, p_{x}\}_{CK} = 1$
        \\
        $\{x', p_{y'}\}_{B} = 0$ & $\{x, p_{y}\}_{CK} = 0$
        \\
        $\{y', p_{x'}\}_{B} = 0$ & $\{y, p_{x}\}_{CK} = 0$
        \\
        $\{y', p_{y'}\}_{B} = 1$ & $\{y, p_{y}\}_{CK} = 1$
        \\
        $\{p_{x'}, p_{y'}\}_{B} = 0$ & $\{p_{x}, p_{y}\}_{CK} = 0$
        \\
        \hline
    \end{tabular}
    \caption{Poisson-bracket preservation under the CK--Bateman transformation.}
    \label{table:canonical_Poisson_brackets}
\end{table}
The brackets obtained by reversing the order follow from antisymmetry, and the Poisson bracket of each variable with itself vanishes. The results in Table~\ref{table:canonical_Poisson_brackets} confirm that the transformation preserves the canonical Poisson-bracket algebra in both directions. The coordinate transformation is invertible because its Jacobian determinant satisfies
$\det[\partial(x, y)/\partial(x', y')] = -1$. 

For the generating function induced by the point transformation in Eq.~\eqref{F2P_final}, the associated boundary function is obtained on the transformation graph from
\begin{equation*}
    F_{1} = F_{2} - x p_{x} - y p_{y},
\end{equation*}
which gives
\begin{equation}
    F_{1} = -\frac{m\gamma}{4} \left(e^{\gamma t} (x')^2 + e^{-\gamma t} (y')^2\right).
    \label{F1P}
\end{equation}
At the classical level, Eqs.~\eqref{coordinate_transformation_x}--\eqref{momentum_transformation_yprime} establish an invertible canonical transformation induced by a time-dependent point transformation between the doubled CK and Bateman systems. Guerrero et al.~\cite{Guerrero2011} obtain a connection by a different route, extending the CK symmetry algebra and recovering the CK model through time-dependent constraints. In the present construction, $H_{CK}$ and $H_{B}$ are canonically related even though $H_{CK}$ is explicitly time dependent, whereas $H_{B}$ is autonomous and conserved.

The transformation relates the complete doubled CK and Bateman systems. It does not imply the direct identifications $x = x'$ and $y = y'$, because each Bateman coordinate is a time-dependent linear combination of the two CK coordinates. The initial conditions of both CK coordinates must therefore be included when the damped motion is expressed in the Bateman variables.
\section{Bateman's Dual-System Approach to a Cosmological Scalar Field}
\label{sec:bateman_cosmological_scalar_field}
The classical Bateman and Caldirola--Kanai formulations developed in the preceding section are now extended to a homogeneous scalar field in a prescribed FLRW background. The discussion focuses on the modifications produced by the time-dependent damping coefficient $3H(t)$. The scale factor $a(t)$ is assumed to be three times continuously differentiable on the time interval considered, while the potential $V(\phi)$ is assumed to be twice continuously differentiable on the relevant field interval.

The potential $V(\phi)$ is retained in a general form during the multiplier construction. The free massive potential is selected subsequently to obtain the linear equations and quadratic Hamiltonians used in the explicit canonical comparison.

\subsection{Homogeneous Scalar-Field Dynamics in an FLRW Background}
\label{subsec:homogeneous_scalar_field_flrw}

Consider a real, minimally coupled canonical scalar field $\phi$ with a potential $V(\phi)$ in a four-dimensional curved spacetime. With the metric signature $(+, -, -, -)$ and natural units $c = \hbar = 1$, the scalar-field matter action is given by~\cite{Copeland2006, Harko2023}
\begin{equation}
    S_{\phi} = \int d^{4}x\,\sqrt{-g}\left[\frac{1}{2}g^{\mu\nu}\partial_{\mu}\phi\,\partial_{\nu}\phi - V(\phi)\right]. \label{action_ideal_scalar_field}
\end{equation}
In Eq.~\eqref{action_ideal_scalar_field}, $g_{\mu\nu}$ denotes the spacetime metric, $g \equiv \det(g_{\mu\nu})$, and $V(\phi)$ is the scalar-field potential. For a scalar field, the covariant derivative reduces to the ordinary derivative, $\nabla_{\mu} \phi = \partial_{\mu} \phi$. The Lagrangian density is defined by
\begin{equation}
    \mathcal{L}_{\phi} = \sqrt{-g}\left[\frac{1}{2}g^{\mu\nu}\partial_{\mu}\phi\,\partial_{\nu}\phi - V(\phi)\right]. \label{scalar_Lagrangian_density}
\end{equation}
The field Euler--Lagrange equation is
\begin{equation}
    \partial_{\mu}\left[\frac{\partial\mathcal{L}_{\phi}}{\partial(\partial_{\mu}\phi)}\right] - \frac{\partial\mathcal{L}_{\phi}}{\partial\phi} = 0. \label{euler_lagrange_scalar_field_covariant}
\end{equation}
This is the field-theoretic form of the Euler--Lagrange equation used in the classical analysis. Substitution of Eq.~\eqref{scalar_Lagrangian_density} into Eq.~\eqref{euler_lagrange_scalar_field_covariant} yields
\begin{equation}
    \frac{1}{\sqrt{-g}} \partial_{\mu}\left(\sqrt{-g}\,g^{\mu\nu}\partial_{\nu}\phi\right) + V'(\phi) = 0, \qquad V'(\phi) \equiv \frac{dV(\phi)}{d\phi}. \label{covariant_KG_phi_expanded}
\end{equation}
The first term in Eq.~\eqref{covariant_KG_phi_expanded} is the covariant d'Alembertian acting on the scalar field,
\begin{equation}
    \Box\phi \equiv \nabla_{\mu}\nabla^{\mu}\phi = \frac{1}{\sqrt{-g}}\partial_{\mu}\left(\sqrt{-g}\,g^{\mu\nu}\partial_{\nu}\phi\right). \label{dalembertian_scalar_field}
\end{equation}
Equation~\eqref{covariant_KG_phi_expanded} then takes the manifestly covariant form
\begin{equation}
    \Box\phi + V'(\phi) = 0. \label{covariant_KG_phi}
\end{equation}
This is the covariant Klein--Gordon equation for a canonical scalar field with a general potential.

For a spatially flat Friedmann--Lema\^itre--Robertson--Walker (FLRW) spacetime, the line element is
\begin{equation}
    ds^{2} = dt^{2} - a^{2}(t) \left(dx^{2} + dy^{2} + dz^{2}\right), \label{metric_FLRW_scalar_field}
\end{equation}
where $a(t) > 0$ is the cosmological scale factor. The scale factor is taken to be dimensionless and is normalized by $a(t_{0}) = 1$ at a chosen reference time $t_{0}$. The determinant of the metric, the invariant volume factor, and the Hubble parameter are
\begin{equation}
    g = -a^{6}(t), \qquad \sqrt{-g} = a^{3}(t), \qquad H(t) \equiv \frac{\dot{a}(t)}{a(t)}. \label{FLRW_determinant_Hubble}
\end{equation}
For a spatially homogeneous scalar field,
\begin{equation}
    \phi = \phi(t), \qquad \partial_{i}\phi = 0, \label{homogeneous_scalar_field}
\end{equation}
so all spatial-gradient terms vanish. Substitution of Eqs.~\eqref{metric_FLRW_scalar_field}--\eqref{homogeneous_scalar_field} into the covariant action \eqref{action_ideal_scalar_field} gives
\begin{align}
    S_{\phi} &= \int dt\,d^{3}x\,a^{3}(t)\left[\frac{1}{2}\dot{\phi}^{\,2}-V(\phi)\right], \nonumber 
    \\
    &= \mathcal{V}_{c}\int dt\,a^{3}(t)\left[\frac{1}{2}\dot{\phi}^{\,2}-V(\phi)\right], \label{action_homogeneous_scalar_field}
\end{align}
where $\mathcal{V}_{c} \equiv \int d^{3}x$ denotes a fiducial comoving spatial volume. Dividing the reduced action by $\mathcal{V}_{c}$ gives the action per unit comoving volume and does not change the equation of motion. The corresponding reduced Lagrangian is
\begin{equation}
    L_{\phi} = a^{3}(t) \left[\frac{1}{2}\dot{\phi}^{\,2} - V(\phi)\right]. \label{reduced_lagrangian_phi}
\end{equation}
Applying the Euler--Lagrange equation to Eq.~\eqref{reduced_lagrangian_phi} gives
\begin{equation}
    \frac{1}{a^{3}(t)}\frac{d}{dt}\left[a^{3}(t)\dot{\phi}\right] + V'(\phi) = 0.
\end{equation}
Using $H(t) = \dot{a}(t)/a(t)$, the first term becomes
\begin{equation}
    \frac{1}{a^{3}}\frac{d}{dt}\left(a^{3}\dot{\phi}\right) = \ddot{\phi} + 3H(t)\dot{\phi}. \label{expansion_FLRW_KG_derivative}
\end{equation}
The homogeneous scalar-field equation of motion is
\begin{equation}
    \ddot{\phi} + 3H(t)\dot{\phi} + V'(\phi) = 0. \label{EOM_phi_FLRW}
\end{equation}
Equation~\eqref{EOM_phi_FLRW} is the homogeneous Klein--Gordon equation in a spatially flat FLRW spacetime~\cite{Copeland2006, Harko2023}. The term $3H(t)\dot{\phi}$ originates from the expansion of the cosmological volume element. For $H(t) > 0$, it acts as a damping term in the reduced homogeneous equation, although it represents a geometrical effect of the expanding background rather than dissipation produced by non-gravitational couplings to a thermal environment~\cite{BanerjeeNguyenTanin2026}. Unlike the constant phenomenological coefficient $\gamma$ in Eq.~\eqref{EOM_x}, the effective damping coefficient $3H(t)$ is generally time dependent.

The energy density and isotropic pressure associated with the homogeneous scalar field are~\cite{Copeland2006, Harko2023}
\begin{equation}
    \rho_{\phi} = \frac{1}{2}\dot{\phi}^{\,2} + V(\phi), 
    \qquad 
    P_{\phi} = \frac{1}{2}\dot{\phi}^{\,2} - V(\phi). \label{density_pressure_phi}
\end{equation}
These quantities describe the physical scalar-field sector and should not be identified with the Bateman Hamiltonian constructed below.

Equation~\eqref{EOM_phi_FLRW} has the same formal structure as the damped oscillator equation \eqref{EOM_x}, with $3H(t)$ playing the role of a time-dependent damping coefficient and $V'(\phi)$ replacing the linear restoring force. The correspondence $\gamma \rightarrow 3H(t)$ motivates the Bateman doubling of the homogeneous scalar-field equation.

\subsection{Bateman Construction from the Multiplier Action}
\label{subsec:bateman_multiplier_construction}

The multiplier method introduced in Subsection~\ref{subsec:bateman_dual_system} is applied to Eq.~\eqref{EOM_phi_FLRW} by introducing a homogeneous auxiliary field $\chi(t)$. The scale factor $a(t)$ and the Hubble parameter $H(t)$ remain prescribed and are not varied. The corresponding multiplier action is
\begin{equation}
    S_{M,\mathrm{SF}}[\phi,\chi] \equiv \int_{t_{i}}^{t_{f}} \chi \left[\ddot{\phi} + 3H(t)\dot{\phi} + V'(\phi)\right]dt. \label{action_Bateman_multiplier_cosmological}
\end{equation}
The overall normalization of the multiplier is arbitrary and is fixed here by assigning $\chi$ the same physical dimension as $\phi$. All reduced auxiliary actions below are understood per unit fiducial comoving volume. The functional $S_{M,\mathrm{SF}}$ is an auxiliary variational functional and should not be identified with the physical scalar-field action $S_{\phi}$ in Eq.~\eqref{action_ideal_scalar_field}. Variation of Eq.~\eqref{action_Bateman_multiplier_cosmological} with respect to $\chi$ gives the homogeneous Klein--Gordon equation \eqref{EOM_phi_FLRW}.

Integration of the term $\chi\ddot{\phi}$ by parts gives
\begin{equation}
    S_{M,\mathrm{SF}}[\phi,\chi] = \left[\chi\dot{\phi}\right]_{t_{i}}^{t_{f}} + S_{1,\mathrm{SF}}[\phi,\chi], 
    \qquad 
    S_{1,\mathrm{SF}}[\phi,\chi] \equiv \int_{t_{i}}^{t_{f}} L_{1,\mathrm{SF}}\,dt, \label{relation_multiplier_first_order_cosmological}
\end{equation}
where
\begin{equation}
    L_{1,\mathrm{SF}} = -\dot{\phi}\dot{\chi} + 3H(t)\chi\dot{\phi} + \chi V'(\phi). \label{lagrangian_Bateman_multiplier_cosmological}
\end{equation}
Variation of $L_{1,\mathrm{SF}}$ with respect to $\chi$ reproduces Eq.~\eqref{EOM_phi_FLRW}, while variation with respect to $\phi$ gives the complementary equation
\begin{equation}
    \ddot{\chi} - 3H(t)\dot{\chi} + \left[V''(\phi) - 3\dot{H}(t)\right]\chi = 0. \label{EOM_chi_FLRW_general}
\end{equation}
Following the classical construction, the boundary term is chosen so that the velocity coupling takes the corresponding Bateman form. A first-order Bateman Lagrangian for a general potential is then defined by
\begin{equation}
    L_{B,\mathrm{SF}}^{(V)} \equiv -L_{1,\mathrm{SF}} + \frac{d}{dt}\left[\frac{3H(t)}{2}\phi\chi\right]. 
    \label{relation_Lagrangians_Bateman_cosmological}
\end{equation}
Evaluation of the total derivative gives
\begin{equation}
    L_{B,\mathrm{SF}}^{(V)} = \dot{\phi}\dot{\chi} - \frac{3H(t)}{2} \left(\chi\dot{\phi} - \phi\dot{\chi}\right) + \frac{3\dot{H}(t)}{2}\phi\chi - \chi V'(\phi). \label{lagrangian_Bateman_general_cosmological}
\end{equation}
The corresponding action is
\begin{equation}
    S_{B,\mathrm{SF}}^{(V)}[\phi,\chi] \equiv \int_{t_{i}}^{t_{f}} L_{B,\mathrm{SF}}^{(V)}\,dt. 
    \label{action_Bateman_general_cosmological}
\end{equation}
Combining Eqs.~\eqref{relation_multiplier_first_order_cosmological} and \eqref{relation_Lagrangians_Bateman_cosmological} gives the direct relation
\begin{equation}
    S_{B,\mathrm{SF}}^{(V)}[\phi,\chi] = -S_{1,\mathrm{SF}}[\phi,\chi] + \left[\frac{3H(t)}{2}\phi\chi\right]_{t_{i}}^{t_{f}} = -S_{M,\mathrm{SF}}[\phi,\chi] + \left[\chi\dot{\phi} + \frac{3H(t)}{2}\phi\chi\right]_{t_{i}}^{t_{f}}. \label{relation_actions_Bateman_cosmological}
\end{equation}
The first equality in Eq.~\eqref{relation_actions_Bateman_cosmological} establishes the fixed-endpoint equivalence of the two first-order variational descriptions. The second equality relates the Bateman action directly to the original second-order multiplier action. As in the classical construction, the derivative-dependent part of the boundary term originates from the integration by parts used to remove the second derivative from the multiplier action.

Variation of Eq.~\eqref{action_Bateman_general_cosmological} with respect to $\chi$ gives Eq.~\eqref{EOM_phi_FLRW}, while variation with respect to $\phi$ gives Eq.~\eqref{EOM_chi_FLRW_general}. Thus, the same Bateman action produces the physical equation and its complementary equation for a general twice continuously differentiable potential. The differential operator in Eq.~\eqref{EOM_chi_FLRW_general} is the formal adjoint of the linearized operator associated with Eq.~\eqref{EOM_phi_FLRW}~\cite{Ibragimov2006, Talukdar2020}. In the present case, the coefficient $3H(t)$ changes to $-3H(t)$ in the adjoint equation, while its time dependence produces the additional term $-3\dot{H}(t)\chi$. For an expanding background with $H(t) > 0$, the term $-3H(t)\dot{\chi}$ has the sign of anti-damping. Since $\chi$ is an auxiliary field, this behavior should not be interpreted as an independent physical instability.

For the explicit canonical comparison developed below, the potential is now specialized to the free massive form
\begin{equation}
    V(\phi) = \frac{1}{2}m^{2}\phi^{2}, 
    \qquad 
    V'(\phi) = m^{2}\phi, 
    \qquad 
    V''(\phi) = m^{2}, 
    \label{potential_free_massive_scalar}
\end{equation}
where $m > 0$ denotes the scalar-field mass. An additive constant $V_{0}$ could be included without changing either equation of motion because it does not contribute to $V'(\phi)$ or $V''(\phi)$. It would, however, contribute to the physical scalar-field energy density and act as a vacuum-energy term if the background geometry were determined self-consistently~\cite{Weinberg1989}. The choice $V_{0} = 0$ is adopted here.

The physical and auxiliary equations then reduce to
\begin{align}
    \ddot{\phi} + 3H(t)\dot{\phi} + m^{2}\phi &= 0, 
    \label{EOM_phi_free_cosmological} 
    \\
    \ddot{\chi} - 3H(t)\dot{\chi} + \left[m^{2} - 3\dot{H}(t)\right]\chi &= 0. \label{EOM_chi_free_cosmological}
\end{align}
Equation~\eqref{EOM_phi_free_cosmological} is the standard equation for a homogeneous massive scalar field in an expanding universe~\cite{Turner1983}. The field $\phi$ is the Hubble-damped member of the pair, while $\chi$ is its anti-damped auxiliary counterpart. The combination $m^{2} - 3\dot{H}(t)$ acts as a time-dependent effective frequency squared for the auxiliary field. It need not be positive for an arbitrary prescribed background; when it is negative, the restoring term in the auxiliary equation changes sign. Since $\chi$ is an auxiliary field, this behavior does not by itself represent a physical scalar-field instability.

\subsection{Bateman Hamiltonian and Its Time Evolution}
\label{subsec:bateman_hamiltonian_cosmological}

For the free massive potential in Eq.~\eqref{potential_free_massive_scalar}, the general Bateman Lagrangian \eqref{lagrangian_Bateman_general_cosmological} becomes
\begin{equation}
    L_{B,\mathrm{SF}} = \dot{\phi}\dot{\chi} - \frac{3H(t)}{2} \left(\chi\dot{\phi} - \phi\dot{\chi}\right) + \frac{3\dot{H}(t)}{2} \phi\chi - m^{2}\phi\chi. \label{lagrangian_Bateman_cosmological}
\end{equation}
The kinetic and velocity-coupling terms have the same algebraic structure as those of the classical Bateman Lagrangian in Eq.~\eqref{lagrangian_Bateman}. The term proportional to $\dot{H}(t)$ originates from the time dependence of the cosmological damping coefficient. When $H(t)$ is constant, the correspondence with the classical system follows from the identifications $x\leftrightarrow\phi$, $y \leftrightarrow \chi$, $\gamma \leftrightarrow 3H$, and $\omega \leftrightarrow m$, with the classical oscillator mass normalized to unity.

The canonical momenta obtained from Eq.~\eqref{lagrangian_Bateman_cosmological} are
\begin{align}
    p_{\phi} \equiv \frac{\partial L_{B,\mathrm{SF}}}{\partial\dot{\phi}} = \dot{\chi}-\frac{3H(t)}{2}\chi
    \quad &\Rightarrow \quad
    \dot{\chi} = p_{\phi} + \frac{3H(t)}{2}\chi, \label{momentum_Bateman_phi_cosmological}
    \\
    p_{\chi} \equiv \frac{\partial L_{B,\mathrm{SF}}}{\partial\dot{\chi}} = \dot{\phi}+\frac{3H(t)}{2}\phi
    \quad &\Rightarrow \quad
    \dot{\phi} = p_{\chi} - \frac{3H(t)}{2}\phi. \label{momentum_Bateman_chi_cosmological}
\end{align}
The Legendre transform of $L_{B,\mathrm{SF}}$ gives the Bateman scalar-field Hamiltonian,
\begin{equation}
    H_{B,\mathrm{SF}} = \dot{\phi}p_{\phi} + \dot{\chi}p_{\chi} - L_{B,\mathrm{SF}} = \dot{\phi}\dot{\chi} + \left[m^{2} - \frac{3\dot{H}(t)}{2}\right]\phi\chi. \label{hamiltonian_Bateman_velocity_cosmological}
\end{equation}
Substitution of Eqs.~\eqref{momentum_Bateman_phi_cosmological} and \eqref{momentum_Bateman_chi_cosmological} into Eq.~\eqref{hamiltonian_Bateman_velocity_cosmological} gives the phase-space Hamiltonian
\begin{equation}
    H_{B,\mathrm{SF}} = p_{\phi}p_{\chi} + \frac{3H(t)}{2}\left(\chi p_{\chi} - \phi p_{\phi}\right) + M_{\mathrm{eff}}^{2}(t)\phi\chi, \label{hamiltonian_Bateman_cosmological}
\end{equation}
where
\begin{equation}
    M_{\mathrm{eff}}^{2}(t) \equiv m^{2} - \frac{9H^{2}(t)}{4} - \frac{3\dot{H}(t)}{2}. \label{effective_mass_Bateman_cosmological}
\end{equation}
As in the classical Bateman system, $H_{B,\mathrm{SF}}$ is not positive definite. It is the Hamiltonian of the enlarged system involving both the physical field $\phi$ and the auxiliary field $\chi$, and should not be identified with the physical scalar-field energy density $\rho_{\phi}$ in Eq.~\eqref{density_pressure_phi}. In contrast to the autonomous Hamiltonian in Eq.~\eqref{hamiltonian_Bateman}, the Bateman scalar-field Hamiltonian generally depends explicitly on time through $H(t)$ and $\dot{H}(t)$.

To analyze the time evolution of $H_{B,\mathrm{SF}}$ in a form parallel to the classical Bateman system, the normalized rotated variables are defined by
\begin{align}
    u &= \frac{\phi+\chi}{\sqrt{2}}, \label{variable_u_cosmological}
    \\
    v &= \frac{\phi-\chi}{\sqrt{2}}. \label{variable_v_cosmological}
\end{align}
The inverse relations are $\phi = (u + v)/\sqrt{2}$ and $\chi = (u - v)/\sqrt{2}$. Adding Eqs.~\eqref{EOM_phi_free_cosmological} and \eqref{EOM_chi_free_cosmological}, and rewriting the result in terms of $u$ and $v$, gives
\begin{equation}
    \ddot{u} + \left[m^{2} - \frac{3\dot{H}(t)}{2}\right]u = -3H(t)\dot{v} - \frac{3\dot{H}(t)}{2}v. \label{EOM_u_cosmological}
\end{equation}
Multiplying Eq.~\eqref{EOM_u_cosmological} by $\dot{u}$ gives the time rate of change of the mode-energy function $E_{u}$:
\begin{equation}
    \frac{dE_{u}}{dt} \equiv \frac{d}{dt}\left\{\frac{1}{2}\dot{u}^{2} + \frac{1}{2}\left[m^{2} - \frac{3\dot{H}(t)}{2}\right]u^{2}\right\} = -3H(t)\dot{u}\dot{v} - \frac{3\dot{H}(t)}{2}v\dot{u} - \frac{3\ddot{H}(t)}{4}u^{2}. 
    \label{dE_u_cosmological}
\end{equation}
Similarly, subtracting Eq.~\eqref{EOM_chi_free_cosmological} from Eq.~\eqref{EOM_phi_free_cosmological} gives
\begin{equation}
    \ddot{v} + \left[m^{2} - \frac{3\dot{H}(t)}{2}\right]v = -3H(t)\dot{u} - \frac{3\dot{H}(t)}{2}u. 
    \label{EOM_v_cosmological}
\end{equation}
Multiplying Eq.~\eqref{EOM_v_cosmological} by $\dot{v}$ gives the time rate of change of $E_{v}$:
\begin{equation}
    \frac{dE_{v}}{dt} \equiv \frac{d}{dt}\left\{\frac{1}{2}\dot{v}^{2} + \frac{1}{2}\left[m^{2} - \frac{3\dot{H}(t)}{2}\right]v^{2}\right\} = -3H(t)\dot{u}\dot{v} - \frac{3\dot{H}(t)}{2}u\dot{v} - \frac{3\ddot{H}(t)}{4}v^{2}. \label{dE_v_cosmological}
\end{equation}
Unlike the corresponding energy rates in the classical Bateman system, Eqs.~\eqref{dE_u_cosmological} and \eqref{dE_v_cosmological} are not generally equal. Their common contribution $-3H(t)\dot{u}\dot{v}$ arises from the damping and anti-damping terms, whereas the additional contributions involving $\dot{H}(t)$ and $\ddot{H}(t)$ result from the time dependence of the cosmological damping coefficient.

Using Eqs.~\eqref{variable_u_cosmological} and \eqref{variable_v_cosmological}, the velocity-space Hamiltonian in Eq.~\eqref{hamiltonian_Bateman_velocity_cosmological} becomes
\begin{align}
    H_{B,\mathrm{SF}} &= \frac{1}{2}\left(\dot{u}^{2} - \dot{v}^{2}\right) + \frac{1}{2}\left[m^{2} - \frac{3\dot{H}(t)}{2}\right] \left(u^{2} - v^{2}\right), \nonumber
    \\
    &= E_{u} - E_{v}. \label{hamiltonian_Bateman_uv_cosmological}
\end{align}
This difference form makes the indefinite character of $H_{B,\mathrm{SF}}$ explicit; the quantities $E_{u}$ and $E_{v}$ are mode-energy functions of the enlarged system rather than independently conserved physical energies. 

The difference between Eqs.~\eqref{dE_u_cosmological} and \eqref{dE_v_cosmological} consequently gives
\begin{equation}
    \frac{dH_{B,\mathrm{SF}}}{dt} \equiv \frac{dE_{u}}{dt} - \frac{dE_{v}}{dt} = \frac{3\dot{H}(t)}{2}\left(u\dot{v} - v\dot{u}\right) - \frac{3\ddot{H}(t)}{4}\left(u^{2} - v^{2}\right).
    \label{Hamiltonian_time_evolution_uv_cosmological}
\end{equation}
The rotated variables satisfy
\begin{equation}
    u\dot{v} - v\dot{u} = \chi\dot{\phi} - \phi\dot{\chi}, 
    \qquad 
    u^{2} - v^{2} = 2\phi\chi. 
    \label{relations_uv_phi_chi_cosmological}
\end{equation}
Equation~\eqref{Hamiltonian_time_evolution_uv_cosmological} may thus be written in terms of the original fields as
\begin{equation}
    \frac{dH_{B,\mathrm{SF}}}{dt} = \frac{3}{2}\dot{H}(t)\left(\chi\dot{\phi} - \phi\dot{\chi}\right) - \frac{3}{2}\ddot{H}(t)\phi\chi. \label{Hamiltonian_time_evolution_cosmological}
\end{equation}
For a constant Hubble parameter, $\dot{H}(t) = \ddot{H}(t) = 0$, and Eq.~\eqref{Hamiltonian_time_evolution_cosmological} shows directly that $H_{B,\mathrm{SF}}$ is conserved. Equation~\eqref{Hamiltonian_time_evolution_cosmological} can also be obtained by differentiating Eq.~\eqref{hamiltonian_Bateman_velocity_cosmological} along a solution and using Eqs.~\eqref{EOM_phi_free_cosmological} and \eqref{EOM_chi_free_cosmological}. For a nonconstant prescribed background, conservation is not generic and holds along a particular trajectory if and only if
\begin{equation}
    \dot{H}(t)\left(\chi\dot{\phi} - \phi\dot{\chi}\right) - \ddot{H}(t)\phi\chi = 0. \label{conservation_condition_Bateman_cosmological}
\end{equation}
For a general prescribed FLRW background, the terms involving $\dot{H}(t)$ and $\ddot{H}(t)$ make the two mode-energy rates unequal. Equation~\eqref{hamiltonian_Bateman_uv_cosmological}, together with Eqs.~\eqref{dE_u_cosmological} and \eqref{dE_v_cosmological}, shows that conservation of $H_{B,\mathrm{SF}}$ is equivalent to equality of the two mode-energy rates. In the classical Bateman system with constant damping, this equality holds identically.

On an interval where $\dot{H}(t)\phi\chi \neq 0$, the conservation condition may be written as
\begin{equation}
    \frac{\chi\dot{\phi} - \phi\dot{\chi}}{\phi\chi} = \frac{\ddot{H}(t)}{\dot{H}(t)},
    \qquad\text{or equivalently}\qquad
    \frac{d}{dt}\ln\left|\frac{\phi}{\chi}\right| = \frac{d}{dt}\ln|\dot{H}(t)|. 
    \label{relative_evolution_conservation_cosmological}
\end{equation}
Integration gives
\begin{equation}
    \frac{\phi}{\chi} = C\dot{H}(t), \label{field_ratio_conservation_cosmological}
\end{equation}
where $C \neq 0$ is constant on the interval considered. Conservation of $H_{B,\mathrm{SF}}$ therefore links the relative evolution of the physical and auxiliary fields to the time dependence of the prescribed background. Since $\phi$ and $\chi$ must also satisfy Eqs.~\eqref{EOM_phi_free_cosmological} and \eqref{EOM_chi_free_cosmological}, Eq.~\eqref{field_ratio_conservation_cosmological} is a compatibility condition on the background and the corresponding field trajectory, rather than an additional equation of motion. At points where $\dot{H}(t)$, $\phi$, or $\chi$ vanishes, conservation must be examined using Eq.~\eqref{conservation_condition_Bateman_cosmological} directly instead of the divided form.

\subsubsection{Power-Law FLRW Background}
\label{subsubsec:power_law_flrw_background}

Consider the prescribed power-law scale factor
\begin{equation}
    a(t) = \left(\frac{t}{t_{0}}\right)^{p}, 
    \qquad 
    H(t) = \frac{p}{t}, 
    \qquad 
    \dot{H}(t) = -\frac{p}{t^{2}}, 
    \qquad
    t > 0, 
    \label{power_law_background}
\end{equation}
where $p > 0$ and $t_{0} > 0$. Equation~\eqref{field_ratio_conservation_cosmological} then gives, after absorbing the constant factor $-1/(Cp)$ into a new constant $K \neq 0$,
\begin{equation}
    \chi(t) = Kt^{2}\phi(t). 
    \label{power_law_conservation_ansatz}
\end{equation}
Substituting Eq.~\eqref{power_law_conservation_ansatz} into the auxiliary equation \eqref{EOM_chi_free_cosmological} and eliminating $\ddot{\phi}$ using the physical equation \eqref{EOM_phi_free_cosmological} gives
\begin{equation}
    (2 - 3p)\left[\phi(t) + 2t\dot{\phi}(t)\right] = 0. \label{power_law_conservation_compatibility}
\end{equation}
For $m > 0$, the second factor in Eq.~\eqref{power_law_conservation_compatibility} cannot vanish identically for a nonzero solution on an interval. Indeed, it would imply $\phi \propto t^{-1/2}$, which is incompatible with Eq.~\eqref{EOM_phi_free_cosmological} for a massive field. A nontrivial correlated family therefore requires
\begin{equation}
    p = \frac{2}{3}. 
    \label{matter_power_law_special}
\end{equation}
The value $p = 2/3$ coincides with the standard matter-dominated FLRW exponent~\cite{Copeland2006}, although the background remains prescribed and no matter-dominated dynamics is derived here. In particular, this value does not make the Hubble parameter constant, since $H(t) = 2/(3t)$. Instead, for each solution $\phi$ of the physical equation and each constant $K \neq 0$, the correlated auxiliary field $\chi = Kt^{2}\phi$ satisfies Eq.~\eqref{EOM_chi_free_cosmological} and fulfills the conservation condition \eqref{conservation_condition_Bateman_cosmological}.

To evaluate the corresponding conserved Hamiltonian, define
\begin{equation}
    q(t) \equiv t\phi(t). 
    \label{q_matter_power_law_definition}
\end{equation}
For $p = 2/3$, Eq.~\eqref{EOM_phi_free_cosmological} reduces to
\begin{equation}
    \ddot{q}(t) + m^{2}q(t) = 0. 
    \label{q_matter_power_law_equation}
\end{equation}
Substitution of $\phi = q/t$ and $\chi = Ktq$ into the velocity-space Hamiltonian \eqref{hamiltonian_Bateman_velocity_cosmological} gives
\begin{equation}
    H_{B,\mathrm{SF}} = K\left[\dot{q}^{\,2}(t) + m^{2}q^{2}(t)\right]. \label{HB_matter_power_law_conserved}
\end{equation}
Equations~\eqref{q_matter_power_law_equation} and \eqref{HB_matter_power_law_conserved} show directly that $H_{B,\mathrm{SF}}$ is constant. The power-law background thus gives a case in which $H(t)$ remains time dependent while $H_{B,\mathrm{SF}}$ is conserved along a correlated family of physical and auxiliary trajectories.

The preceding conservation results apply only to the reduced Bateman field system in which $H(t)$ is treated as a prescribed background function. For comparison, an open-system description coupled to dynamical gravity must also include the energy--momentum tensor of the environment because gravity couples to all degrees of freedom~\cite{LauNishiiNoumi2025}. If the scale factor is promoted to a dynamical variable, $H_{B,\mathrm{SF}}$ no longer represents the Hamiltonian of the complete system, and conservation must be reexamined in the enlarged scalar--gravity phase space.
\section{Doubled Caldirola--Kanai Formulation of the Cosmological Scalar-Field System}
\label{sec:doubled_ck_cosmological_scalar_field}
The preceding section established a Bateman Lagrangian for a general scalar potential and then specialized to a free massive field to construct the Hamiltonian for the homogeneous pair $(\phi,\chi)$. For the free massive specialization adopted below, the physical and auxiliary equations are dynamically decoupled, although their Bateman Lagrangian remains bilinear and provides a coupled canonical description. The Caldirola--Kanai formulation gives an alternative description in terms of separate time-dependent Lagrangians for the damped and anti-damped sectors.

To distinguish the CK canonical variables from the Bateman variables $(\phi, \chi, p_{\phi}, p_{\chi})$, the CK coordinates are denoted by $(\phi', \chi')$ and their conjugate momenta by $(p_{\phi'}, p_{\chi'})$. The primes distinguish the two canonical descriptions and do not denote differentiation. The free massive potential in Eq.~\eqref{potential_free_massive_scalar} is adopted because it gives linear physical and auxiliary CK sectors and a quadratic doubled Hamiltonian. For a general nonlinear potential, the coefficient $V''(\phi)$ in the auxiliary equation depends on the physical trajectory; hence the uncoupled doubled CK construction used below is restricted to the quadratic case. As before, $a(t)$ and $H(t)$ remain prescribed background functions.

\subsection{Caldirola--Kanai Lagrangians for the Damped and Anti-Damped Sectors}
\label{subsec:ck_lagrangians_cosmological}
For the damped CK coordinate $\phi'$, Eq.~\eqref{EOM_phi_free_cosmological} follows from the reduced scalar-field Lagrangian in Eq.~\eqref{reduced_lagrangian_phi}, after using the quadratic potential in Eq.~\eqref{potential_free_massive_scalar} and replacing $\phi$ by $\phi'$:
\begin{align}
    L_{\phi'}(\phi', \dot{\phi}', t) &= a^{3}(t)\left[\frac{1}{2}\left(\dot{\phi}'\right)^{2} - \frac{1}{2}m^{2}\left(\phi'\right)^{2}\right], \nonumber
    \\
    &= \frac{1}{2}a^{3}(t)\left(\dot{\phi}'\right)^{2} - \frac{1}{2}a^{3}(t)m^{2}\left(\phi'\right)^{2}. 
    \label{lagrangian_CK_phi_cosmological}
\end{align}
Applying the Euler--Lagrange equation to $L_{\phi'}$ gives
\begin{equation*}
    \frac{d}{dt}\left(\frac{\partial L_{\phi'}}{\partial\dot{\phi}'}\right) - \frac{\partial L_{\phi'}}{\partial\phi'} = a^{3}(t)\left[\ddot{\phi}' + 3H(t)\dot{\phi}' + m^{2}\phi'\right] = 0.
\end{equation*}
The generalized CK multiplier is consistent with the standard Lagrangian class for second-order equations with linear time-dependent damping~\cite{Musielak2008}. Since $a(t_{0}) = 1$ and $\frac{d}{dt}\ln a^{3}(t) = 3H(t)$, setting $\gamma(t) = 3H(t)$ in the normalized multiplier $g(t) = \exp[\int_{t_{0}}^{t}\gamma(\tau)\,d\tau]$~\cite{Cha2015} gives $g(t) = a^{3}(t)$. For constant $\gamma$ and $t_{0} = 0$, this multiplier reduces to the factor $e^{\gamma t}$ in Eq.~\eqref{lagrangian_CK_xprime}.

For the anti-damped CK coordinate $\chi'$, a CK-type Lagrangian is required to reproduce Eq.~\eqref{EOM_chi_free_cosmological} with $\chi$ replaced by $\chi'$. Introducing a nonvanishing time-dependent multiplier $f_{\chi'}(t)$ gives
\begin{equation}
    L_{\chi'} = f_{\chi'}(t)\left\{\frac{1}{2}\left(\dot{\chi}'\right)^{2} - \frac{1}{2}\left[m^{2} - 3\dot{H}(t)\right]\left(\chi'\right)^{2}\right\}. \label{lagrangian_CK_chi_ansatz_cosmological}
\end{equation}
The Euler--Lagrange equation is
\begin{equation}
    \ddot{\chi}' + \frac{\dot{f}_{\chi'}(t)}{f_{\chi'}(t)}\dot{\chi}' + \left[m^{2} - 3\dot{H}(t)\right]\chi' = 0.
    \label{EOM_CK_chi_integrating_factor}
\end{equation}
Matching Eq.~\eqref{EOM_CK_chi_integrating_factor} with Eq.~\eqref{EOM_chi_free_cosmological}, after replacing $\chi$ by $\chi'$, requires
\begin{equation}
    \frac{\dot{f}_{\chi'}(t)}{f_{\chi'}(t)} = -3H(t). \label{integrating_factor_CK_chi_condition}
\end{equation}
Using $H(t) = \dot{a}(t)/a(t)$, integration gives
\begin{equation}
    f_{\chi'}(t) = C_{\chi}\exp\left[-3\int_{t_{0}}^{t}H(\tau)\,d\tau\right] = C_{\chi}a^{-3}(t),
    \label{integrating_factor_CK_chi_solution}
\end{equation}
where $C_{\chi} \neq 0$ is constant. This overall factor does not affect the Euler--Lagrange equation, so the normalization $C_{\chi} = 1$ is adopted. The auxiliary CK Lagrangian is then
\begin{align}
    L_{\chi'}(\chi', \dot{\chi}', t) &= a^{-3}(t)\left\{\frac{1}{2}\left(\dot{\chi}'\right)^{2} - \frac{1}{2}\left[m^{2} - 3\dot{H}(t)\right]\left(\chi'\right)^{2}\right\}, \nonumber
    \\
    &= \frac{1}{2}a^{-3}(t)\left(\dot{\chi}'\right)^{2} - \frac{1}{2}a^{-3}(t)\left[m^{2} - 3\dot{H}(t)\right]\left(\chi'\right)^{2}. 
    \label{lagrangian_CK_chi_cosmological}
\end{align}
With this choice of integrating factor, the Euler--Lagrange equation of $L_{\chi'}$ reproduces Eq.~\eqref{EOM_chi_free_cosmological} under the relabeling $\chi \rightarrow \chi'$. The factors $a^{3}(t)$ and $a^{-3}(t)$ in Eqs.~\eqref{lagrangian_CK_phi_cosmological} and \eqref{lagrangian_CK_chi_cosmological} generate the damping and anti-damping terms, respectively. Because multiplication by a nonzero constant and addition of a total time derivative do not change the Euler--Lagrange equations, the individual CK Lagrangians do not have unique normalizations. The choices above fix their relative scaling in the doubled CK system used below.

\subsection{Doubled CK Lagrangian and Hamiltonian}
\label{subsec:ck_hamiltonian_cosmological}
The damped and anti-damped CK sectors are combined with a relative minus sign, following the doubled construction used in the classical system~\cite{Cariglia2016}:
\begin{equation}
    L_{CK, \mathrm{SF}} \equiv L_{\phi'} - L_{\chi'}. \label{lagrangian_CK_definition_cosmological}
\end{equation}
Substitution of Eqs.~\eqref{lagrangian_CK_phi_cosmological} and \eqref{lagrangian_CK_chi_cosmological} gives
\begin{equation}
    L_{CK, \mathrm{SF}} = \frac{1}{2}a^{3}(t)\left[\left(\dot{\phi}'\right)^{2} - m^{2}\left(\phi'\right)^{2}\right] - \frac{1}{2}a^{-3}(t)\left\{\left(\dot{\chi}'\right)^{2} - \left[m^{2} - 3\dot{H}(t)\right]\left(\chi'\right)^{2}\right\}. 
    \label{lagrangian_CK_total_cosmological}
\end{equation}
The relative minus sign leaves the Euler--Lagrange equation of each uncoupled sector unchanged and produces the same indefinite difference structure as the classical doubled CK system.

The canonical momenta obtained from Eq.~\eqref{lagrangian_CK_total_cosmological} are
\begin{alignat}{2}
    p_{\phi'} &\equiv \frac{\partial L_{CK,\mathrm{SF}}}{\partial\dot{\phi}'} = a^{3}(t)\dot{\phi}'
    \quad &\Rightarrow\quad
    \dot{\phi}' &= a^{-3}(t)p_{\phi'},
    \label{momentum_CK_phi_cosmological}
    \\
    p_{\chi'} &\equiv\frac{\partial L_{CK,\mathrm{SF}}}{\partial\dot{\chi}'} = -a^{-3}(t)\dot{\chi}'
    \quad &\Rightarrow\quad
    \dot{\chi}' &= -a^{3}(t)p_{\chi'}.
    \label{momentum_CK_chi_cosmological}
\end{alignat}
The minus sign in $p_{\chi'}$ follows from the negative kinetic term of the $\chi'$ sector in the doubled Lagrangian. The Legendre transform gives
\begin{align}
    H_{CK,\mathrm{SF}} &= \dot{\phi}'p_{\phi'} + \dot{\chi}'p_{\chi'} - L_{CK,\mathrm{SF}}, \nonumber
    \\
    &=\frac{1}{2}a^{-3}(t)p_{\phi'}^{2} - \frac{1}{2}a^{3}(t)p_{\chi'}^{2} + \frac{1}{2}a^{3}(t)m^{2}\left(\phi'\right)^{2} - \frac{1}{2}a^{-3}(t)\left[m^{2} - 3\dot{H}(t)\right]\left(\chi'\right)^{2}. 
    \label{hamiltonian_CK_total_cosmological}
\end{align}
The Hamiltonian may also be written in the difference form
\begin{equation}
    H_{CK, \mathrm{SF}} = H_{\phi'} - H_{\chi'}, 
    \label{hamiltonian_CK_difference_cosmological}
\end{equation}
where
\begin{align}
    H_{\phi'} &= \frac{1}{2}a^{-3}(t)p_{\phi'}^{2} + \frac{1}{2}a^{3}(t)m^{2}\left(\phi'\right)^{2}, 
    \label{hamiltonian_CK_phi_cosmological}
    \\
    H_{\chi'} &= \frac{1}{2}a^{3}(t)p_{\chi'}^{2} + \frac{1}{2}a^{-3}(t)\left[m^{2} - 3\dot{H}(t)\right]\left(\chi'\right)^{2}. 
    \label{hamiltonian_CK_chi_cosmological}
\end{align}
The variables $\phi'$ and $\chi'$ form two uncoupled CK sectors that reproduce Eqs.~\eqref{EOM_phi_free_cosmological} and \eqref{EOM_chi_free_cosmological} under the relabelings $\phi \rightarrow \phi'$ and $\chi \rightarrow \chi'$, respectively. The doubled CK Hamiltonian depends explicitly on time through $a(t)$ and $\dot{H}(t)$ and is not generally conserved. Its relation to the coupled Bateman Hamiltonian will be established through a time-dependent canonical transformation in the following section.
\section{Canonical Transformation between the Doubled CK and Bateman Scalar-Field Systems}
\label{sec:cosmological_ck_bateman_transformation}
The CK pair $(\phi', \chi')$ constructed in the preceding section reproduces the damped and anti-damped equations of the Bateman pair $(\phi, \chi)$ under the corresponding relabelings, but the two formulations use different canonical variables. The classical transformation in Subsection~\ref{subsec:classical_ck_bateman_transformation} suggests the replacements $e^{\pm\gamma t} \rightarrow a^{\pm3}(t)$ and $\gamma \rightarrow 3H(t)$, with the classical oscillator mass normalized to unity. These replacements alone do not establish canonical equivalence because the time derivative of the cosmological generating function also produces terms proportional to $\dot{H}(t)$. After expressing the Bateman variables in terms of the CK variables, the coefficients are determined by imposing
\begin{equation}
    L_{CK, \mathrm{SF}} = L_{B,\mathrm{SF}} + \frac{dF_{1,\mathrm{SF}}}{dt}, \label{relation_Lagrangians_Bateman_CK_cosmological}
\end{equation}
where $F_{1,\mathrm{SF}}$ is the boundary function associated with the transformation. Using the Legendre-transform identities for the cosmological CK and Bateman systems, Eq.~\eqref{relation_Lagrangians_Bateman_CK_cosmological} takes the first-order form
\begin{equation}
    p_{\phi'}\dot{\phi}' + p_{\chi'}\dot{\chi}' - H_{CK,\mathrm{SF}} = p_{\phi}\dot{\phi} + p_{\chi}\dot{\chi} - H_{B,\mathrm{SF}} + \frac{dF_{1,\mathrm{SF}}}{dt}. \label{relation_Hamiltonian_Bateman_CK_cosmological}
\end{equation}
The type-2 generating-function formalism derived in Subsection~\ref{subsec:classical_ck_bateman_transformation} applies directly to the cosmological system. A generating function of the form
\begin{equation}
    F_{2,\mathrm{SF}} = A(\phi',\chi',t)p_{\phi} + B(\phi',\chi',t)p_{\chi} + G(\phi',\chi',t)
    \label{F2_ansatz_cosmological}
\end{equation}
is therefore considered. Since the transformation acts as a point transformation on the coordinates, $F_{2,\mathrm{SF}}$ is linear in the Bateman momenta. The corresponding canonical conditions are
\begin{equation}
\begin{aligned}
    p_{\phi'} &= \frac{\partial F_{2,\mathrm{SF}}}{\partial\phi'},
    &\qquad
    p_{\chi'} &= \frac{\partial F_{2,\mathrm{SF}}}{\partial\chi'},
    \\
    \phi &= \frac{\partial F_{2,\mathrm{SF}}}{\partial p_{\phi}},
    &\qquad
    \chi &= \frac{\partial F_{2,\mathrm{SF}}}{\partial p_{\chi}}.
\end{aligned}
    \label{canonical_conditions_cosmological}
\end{equation}
The transformed Hamiltonians must also satisfy
\begin{equation}
    H_{B,\mathrm{SF}} = H_{CK,\mathrm{SF}} + \frac{\partial F_{2,\mathrm{SF}}}{\partial t}.
    \label{relation_Hamiltonians_Bateman_CK_cosmological}
\end{equation}
As in the classical construction, the transformation is restricted to a linear form that leaves the origin fixed. Accordingly, $A$ and $B$ are taken to be linear in $(\phi', \chi')$, while $G$ is homogeneous quadratic.
\begin{align}
    A &= \alpha_{1}(t)\phi' + \beta_{1}(t)\chi',
    \qquad
    B = \alpha_{2}(t)\phi' + \beta_{2}(t)\chi', 
    \label{AB_ansatz_cosmological}
    \\
    G &= \frac{1}{2}g_{11}(t)(\phi')^{2} + g_{12}(t)\phi'\chi' + \frac{1}{2}g_{22}(t)(\chi')^{2}. 
    \label{G_ansatz_cosmological}
\end{align}
The canonical conditions in Eq.~\eqref{canonical_conditions_cosmological} give
\begin{align}
    p_{\phi'} &= \alpha_{1}p_{\phi} + \alpha_{2}p_{\chi} + g_{11}\phi' + g_{12}\chi', 
    \label{momentum_phi_prime_ansatz_cosmological} 
    \\
    p_{\chi'} &= \beta_{1}p_{\phi} + \beta_{2}p_{\chi} + g_{12}\phi' + g_{22}\chi', 
    \label{momentum_chi_prime_ansatz_cosmological} 
    \\
    \phi &= \alpha_{1}\phi' + \beta_{1}\chi', 
    \label{coordinate_phi_ansatz_cosmological}
    \\
    \chi &= \alpha_{2}\phi' + \beta_{2}\chi'. 
    \label{coordinate_chi_ansatz_cosmological}
\end{align}
The explicit time derivative of the generating function is
\begin{align*}
    \frac{\partial F_{2,\mathrm{SF}}}{\partial t} &= \dot{\alpha}_{1}\phi'p_{\phi} + \dot{\beta}_{1}\chi'p_{\phi} + \dot{\alpha}_{2}\phi'p_{\chi} + \dot{\beta}_{2}\chi'p_{\chi} +\frac{1}{2}\dot{g}_{11}(\phi')^{2} + \dot{g}_{12}\phi'\chi' + \frac{1}{2}\dot{g}_{22}(\chi')^{2}.
\end{align*}
Inserting these expressions into Eq.~\eqref{relation_Hamiltonians_Bateman_CK_cosmological} reduces the Hamiltonian relation to coefficient matching among the ten independent monomials
\begin{equation*}
    p_{\phi}^{2}, \quad
    p_{\chi}^{2}, \quad
    p_{\phi}p_{\chi}, \quad
    (\phi')^{2}, \quad
    (\chi')^{2}, \quad
    \phi'\chi', \quad
    p_{\phi}\phi', \quad
    p_{\phi}\chi', \quad
    p_{\chi}\phi', \quad
    p_{\chi}\chi'.
\end{equation*}
Their coefficients must satisfy
\begin{align}
    \frac{a^{-3}(t)}{2}\alpha_{1}^{2} - \frac{a^{3}(t)}{2}\beta_{1}^{2} &= 0, 
    \label{coefficient_pphi2_cosmological} 
    \\
    \frac{a^{-3}(t)}{2}\alpha_{2}^{2} - \frac{a^{3}(t)}{2}\beta_{2}^{2} &= 0, 
    \label{coefficient_pchi2_cosmological} 
    \\
    a^{-3}(t)\alpha_{1}\alpha_{2} - a^{3}(t)\beta_{1}\beta_{2} &= 1, 
    \label{coefficient_pphi_pchi_cosmological} 
    \\
    \frac{a^{-3}(t)}{2}g_{11}^{2} - \frac{a^{3}(t)}{2}g_{12}^{2} + \frac{a^{3}(t)m^{2}}{2} - M_{\mathrm{eff}}^{2}(t)\alpha_{1}\alpha_{2} + \frac{1}{2}\dot{g}_{11} &= 0, 
    \label{coefficient_phi_prime2_cosmological} 
    \\
    \frac{a^{-3}(t)}{2}g_{12}^{2} - \frac{a^{3}(t)}{2}g_{22}^{2} - \frac{a^{-3}(t)}{2}\left[m^{2}-3\dot{H}(t)\right] - M_{\mathrm{eff}}^{2}(t)\beta_{1}\beta_{2} + \frac{1}{2}\dot{g}_{22} &= 0, 
    \label{coefficient_chi_prime2_cosmological} 
    \\
    a^{-3}(t)g_{11}g_{12} - a^{3}(t)g_{12}g_{22} - M_{\mathrm{eff}}^{2}(t) \left(\alpha_{1}\beta_{2}+\alpha_{2}\beta_{1}\right) + \dot{g}_{12} &= 0, 
    \label{coefficient_phi_prime_chi_prime_cosmological} 
    \\
    a^{-3}(t)\alpha_{1}g_{11} - a^{3}(t)\beta_{1}g_{12} + \dot{\alpha}_{1} + \frac{3H(t)}{2}\alpha_{1} &= 0, 
    \label{coefficient_pphi_phi_prime_cosmological} 
    \\
    a^{-3}(t)\alpha_{1}g_{12} - a^{3}(t)\beta_{1}g_{22} + \dot{\beta}_{1} + \frac{3H(t)}{2}\beta_{1} &= 0, 
    \label{coefficient_pphi_chi_prime_cosmological} 
    \\
    a^{-3}(t)\alpha_{2}g_{11} - a^{3}(t)\beta_{2}g_{12} + \dot{\alpha}_{2} - \frac{3H(t)}{2}\alpha_{2} &= 0, 
    \label{coefficient_pchi_phi_prime_cosmological} 
    \\
    a^{-3}(t)\alpha_{2}g_{12} - a^{3}(t)\beta_{2}g_{22} + \dot{\beta}_{2} - \frac{3H(t)}{2}\beta_{2} &= 0. 
    \label{coefficient_pchi_chi_prime_cosmological}
\end{align}
Equations~\eqref{coefficient_pphi2_cosmological} and \eqref{coefficient_pchi2_cosmological} imply
\begin{equation*}
    \beta_{i} = s_{i}a^{-3}(t)\alpha_{i},
    \qquad
    s_{i} \in \{+1, -1\},
    \qquad
    i = 1, 2.
\end{equation*}
The branches with $s_{1} = s_{2}$ eliminate the required $p_{\phi}p_{\chi}$ term and are excluded. The two remaining branches are related by the simultaneous sign reversal $(\chi', p_{\chi'}) \rightarrow (-\chi', -p_{\chi'})$. Choosing $s_{1} = +1$ and $s_{2} = -1$ gives
\begin{equation}
    \beta_{1} = a^{-3}(t)\alpha_{1},
    \qquad
    \beta_{2} = -a^{-3}(t)\alpha_{2}.
    \label{beta_relations_cosmological}
\end{equation}
Equation~\eqref{coefficient_pphi_pchi_cosmological} then fixes the product
\begin{equation}
    \alpha_{1}(t)\alpha_{2}(t) = \frac{1}{2}a^{3}(t).
    \label{alpha_product_cosmological}
\end{equation}
With Eq.~\eqref{beta_relations_cosmological}, the four momentum-coordinate conditions imply
\begin{equation}
    g_{12}(t) = 0, \label{g12_cosmological}
\end{equation}
together with
\begin{equation*}
    \frac{\dot{\alpha}_{2}}{\alpha_{2}} - \frac{\dot{\alpha}_{1}}{\alpha_{1}} = 3H(t).
\end{equation*}
A second relation follows by differentiating Eq.~\eqref{alpha_product_cosmological} with respect to time:
\begin{equation*}
    \frac{\dot{\alpha}_{1}}{\alpha_{1}} + \frac{\dot{\alpha}_{2}}{\alpha_{2}} = 3H(t).
\end{equation*}
Solving the sum and difference of these relations gives
\begin{equation}
    \frac{\dot{\alpha}_{1}}{\alpha_{1}} = 0,
    \qquad
    \frac{\dot{\alpha}_{2}}{\alpha_{2}} = 3H(t).
    \label{alpha_derivatives_cosmological}
\end{equation}
The first relation implies $\alpha_{1} = c \neq 0$, and Eq.~\eqref{alpha_product_cosmological} then fixes
\begin{equation*}
    \alpha_{2} = \frac{a^{3}(t)}{2c}.
\end{equation*}
The constant $c$ represents a reciprocal rescaling of the two sectors. Choosing $c = 1/\sqrt{2}$ gives the normalized coefficients
\begin{equation}
    \alpha_{1} = \frac{1}{\sqrt{2}},
    \qquad
    \alpha_{2} = \frac{a^{3}(t)}{\sqrt{2}},
    \qquad
    \beta_{1} = \frac{a^{-3}(t)}{\sqrt{2}},
    \qquad
    \beta_{2} = -\frac{1}{\sqrt{2}}. 
    \label{linear_coefficients_cosmological}
\end{equation}
The mixed-coefficient equations also determine the diagonal quadratic coefficients:
\begin{equation}
    g_{11}(t) = -\frac{3H(t)}{2}a^{3}(t),
    \qquad
    g_{22}(t) = -\frac{3H(t)}{2}a^{-3}(t). 
    \label{diagonal_coefficients_cosmological}
\end{equation}
With $M_{\mathrm{eff}}^{2}(t)$ defined by Eq.~\eqref{effective_mass_Bateman_cosmological}, the three coordinate-quadratic conditions are satisfied identically and impose no additional restriction on the prescribed FLRW background.
Collecting these coefficients gives the generating function
\begin{align}
    F_{2,\mathrm{SF}} &= \frac{1}{\sqrt{2}} \left[\phi' + a^{-3}(t)\chi'\right]p_{\phi} + \frac{1}{\sqrt{2}} \left[a^{3}(t)\phi' - \chi'\right]p_{\chi} - \frac{3H(t)}{4} \left[a^{3}(t)(\phi')^{2} + a^{-3}(t)(\chi')^{2}\right]. 
    \label{F2_final_cosmological}
\end{align}
The forward phase-space map generated by Eq.~\eqref{F2_final_cosmological} is
\begin{align}
    \phi &= \frac{\phi' + a^{-3}(t)\chi'}{\sqrt{2}}, 
    \label{coordinate_transformation_phi_cosmological} 
    \\
    \chi &= \frac{a^{3}(t)\phi' - \chi'}{\sqrt{2}}, 
    \label{coordinate_transformation_chi_cosmological} 
    \\
    p_{\phi} &= \frac{p_{\phi'} + a^{3}(t)p_{\chi'}}{\sqrt{2}} + \frac{3H(t)}{2\sqrt{2}} \left[a^{3}(t)\phi' + \chi'\right], 
    \label{momentum_transformation_phi_cosmological} 
    \\
    p_{\chi} &= \frac{a^{-3}(t)p_{\phi'} - p_{\chi'}}{\sqrt{2}} + \frac{3H(t)}{2\sqrt{2}} \left[\phi' - a^{-3}(t)\chi'\right]. 
    \label{momentum_transformation_chi_cosmological}
\end{align}
Solving these relations for the primed variables gives the inverse map
\begin{align}
    \phi' &= \frac{\phi + a^{-3}(t)\chi}{\sqrt{2}}, 
    \label{coordinate_transformation_phi_prime_cosmological} 
    \\
    \chi' &= \frac{a^{3}(t)\phi - \chi}{\sqrt{2}}, 
    \label{coordinate_transformation_chi_prime_cosmological} 
    \\
    p_{\phi'} &= \frac{p_{\phi} + a^{3}(t)p_{\chi}}{\sqrt{2}} - \frac{3H(t)}{2\sqrt{2}} \left[a^{3}(t)\phi + \chi\right], 
    \label{momentum_transformation_phi_prime_cosmological} 
    \\
    p_{\chi'} &= \frac{a^{-3}(t)p_{\phi} - p_{\chi}}{\sqrt{2}} - \frac{3H(t)}{2\sqrt{2}} \left[\phi - a^{-3}(t)\chi\right]. 
    \label{momentum_transformation_chi_prime_cosmological}
\end{align}
The generating-function construction already ensures canonicity. For an equal-time check, the Bateman variables in Eqs.~\eqref{coordinate_transformation_phi_cosmological}--\eqref{momentum_transformation_chi_cosmological} are treated as functions of the CK phase-space variables $(\phi', \chi', p_{\phi'}, p_{\chi'})$. The notation $\{\cdot, \cdot\}_{CK,\mathrm{SF}}$ denotes a Poisson bracket evaluated with respect to these primed variables. At fixed $t$, the functions $a^{\pm3}(t)$ and $H(t)$ act as prescribed coefficients, giving
\begin{equation}
    \{\phi,\chi\}_{CK,\mathrm{SF}} = 0, 
    \qquad
    \{\phi,p_{\phi}\}_{CK,\mathrm{SF}} = 1, 
    \qquad
    \{\chi,p_{\chi}\}_{CK,\mathrm{SF}} = 1,
\end{equation}
together with
\begin{equation}
    \{\phi,p_{\chi}\}_{CK,\mathrm{SF}} = \{\chi,p_{\phi}\}_{CK,\mathrm{SF}} = \{p_{\phi},p_{\chi}\}_{CK,\mathrm{SF}} = 0.
\end{equation}
The coordinate Jacobian is
\begin{equation}
    \det\left[\frac{\partial(\phi,\chi)}{\partial(\phi',\chi')}\right] = -1.
\end{equation}
Since this determinant is nonzero, the coordinate map is invertible. The inverse of a canonical transformation is also canonical, so the primed variables preserve their canonical brackets when evaluated with respect to the Bateman phase-space variables.
On the transformation graph, $\phi = A$ and $\chi = B$, so the momentum-dependent terms in $F_{2,\mathrm{SF}} - \phi p_{\phi} - \chi p_{\chi}$ cancel. The boundary function reduces to
\begin{equation}
    F_{1,\mathrm{SF}} = -\frac{3H(t)}{4} \left[a^{3}(t)(\phi')^{2} + a^{-3}(t)(\chi')^{2}\right]. 
    \label{F1_cosmological}
\end{equation}
Preservation of the equal-time Poisson brackets is not sufficient to establish equivalence of the time-dependent Hamiltonians. Equation~\eqref{relation_Hamiltonians_Bateman_CK_cosmological} must also be satisfied. Holding $(\phi', \chi', p_{\phi}, p_{\chi})$ fixed, the required partial derivative is
\begin{align}
    \frac{\partial F_{2,\mathrm{SF}}}{\partial t} &= -\frac{3H(t)}{\sqrt{2}}a^{-3}(t)\chi'p_{\phi} + \frac{3H(t)}{\sqrt{2}}a^{3}(t)\phi'p_{\chi} \nonumber 
    \\
    &\quad - \frac{3}{4}a^{3}(t) \left[\dot{H}(t) + 3H^{2}(t)\right](\phi')^{2} - \frac{3}{4}a^{-3}(t) \left[\dot{H}(t) - 3H^{2}(t)\right](\chi')^{2}. 
    \label{F2_time_derivative_cosmological}
\end{align}
Expressed in the Bateman variables, the combination on the right-hand side of Eq.~\eqref{relation_Hamiltonians_Bateman_CK_cosmological} is
\begin{align}
    H_{CK,\mathrm{SF}} + \frac{\partial F_{2,\mathrm{SF}}}{\partial t} &= p_{\phi}p_{\chi} + \frac{3H(t)}{2} \left(\chi p_{\chi} - \phi p_{\phi}\right) + \left[ m^{2} - \frac{9H^{2}(t)}{4} - \frac{3\dot{H}(t)}{2} \right]\phi\chi, \nonumber 
    \\
    &= H_{B,\mathrm{SF}}.
    \label{Hamiltonian_equivalence_cosmological}
\end{align}
The terms proportional to $\dot{H}(t)$ provide a direct check of the transformation. The auxiliary CK Hamiltonian in Eq.~\eqref{hamiltonian_CK_chi_cosmological} and the time derivative in Eq.~\eqref{F2_time_derivative_cosmological} combine to give the coefficient of $\phi\chi$ in Eq.~\eqref{Hamiltonian_equivalence_cosmological}. Within the linear point-transformation construction, the auxiliary term $-3\dot{H}(t)$ is required for this equality; without it, the coordinate-quadratic terms would not match.

In the static limit $a(t) = 1$ and $H(t) = \dot{H}(t) = 0$, the quadratic part of $F_{2,\mathrm{SF}}$ vanishes. Equations~\eqref{coordinate_transformation_phi_cosmological} and \eqref{coordinate_transformation_chi_cosmological} then reduce to the normalized sum-and-difference transformation obtained from the classical construction in the absence of damping.

Within the linear point-transformation ansatz of Eqs.~\eqref{AB_ansatz_cosmological} and \eqref{G_ansatz_cosmological}, Eqs.~\eqref{coordinate_transformation_phi_cosmological}--\eqref{Hamiltonian_equivalence_cosmological} establish an invertible time-dependent canonical map between the doubled CK and Bateman scalar-field systems. The explicit time dependence explains why canonical equivalence does not require the two Hamiltonian functions to be equal or to possess the same conservation properties.

This equivalence applies only to the complete doubled systems. Since the transformation mixes $\phi'$ and $\chi'$, it implies neither the direct identification $\phi = \phi'$ nor equivalence of the one-field subsystems obtained by discarding the auxiliary variables. Physical quantities defined from $\phi$ must be carried through the complete phase-space map, with the initial data of both sectors transformed together. Throughout the construction, $a(t)$ is prescribed. If the scale factor is promoted to a dynamical variable, its conjugate momentum, the gravitational Hamiltonian, and the associated constraint structure must be included in the enlarged phase space.
\section{Discussion and Conclusions}
\label{sec:discussion_conclusions}

The Bateman formulation separates the equations of motion of the damped and amplified coordinates while coupling these variables in the Lagrangian and Hamiltonian. After the normalized rotation to $u$ and $v$, Eqs.~\eqref{dE_u} and \eqref{dE_v} show that the two mode-energy functions have equal time derivatives, while Eq.~\eqref{dH_B} gives the conservation of their difference. This conserved difference does not represent the sum or difference of the ordinary mechanical energies associated with $x$ and $y$. Consequently, the auxiliary oscillator cannot generally be regarded as an independent reservoir that simply receives the mechanical energy lost by the damped oscillator, in agreement with the discussion of Schuch~\cite{Schuch2015}. For comparison, Hamiltonian reservoir models of linear friction introduce explicit environmental degrees of freedom that exchange energy and momentum with the particle while conserving the total energy and momentum~\cite{BruneauDeBievre2002}. Scalar-field reservoir models make this distinction explicit by coupling the system to environmental Klein--Gordon fields and calculating the associated energy transfer~\cite{KheirandishAmooshahi2006}. Covariant extensions also yield response, noise, and fluctuation--dissipation relations~\cite{RefaeiKheirandish2016}.

The doubled CK and Bateman systems reproduce the same damped and amplified equations but use different canonical variables. The Hamiltonian relation in Eq.~\eqref{relation_Hamiltonians_Bateman_CK} is the one required by a time-dependent canonical transformation; hence the two Hamiltonians need not have the same explicit time dependence or conservation properties. The transformation obtained here retains both sectors and is induced by a linear point transformation. Its configuration-space character makes the time-dependent coordinate rescalings explicit and separates them from the momentum shifts required by Hamiltonian equivalence. This differs from the phase-space transformation of Cariglia et al.~\cite{Cariglia2016}, the symmetry-algebra construction and constraint reduction of Guerrero et al.~\cite{Guerrero2011}, and the indirect connection through expanding coordinates and constraints considered by Schuch et al.~\cite{Schuch2015}.

In the cosmological extension, the time dependence of the damping coefficient produces the term $-3\dot{H}(t)\chi$ in the complementary equation. This term follows directly from the variation of the multiplier action and is required for Hamiltonian equivalence under the canonical transformation constructed here. The multiplier action and its first-order Bateman representative reproduce the physical and complementary equations for a general twice continuously differentiable potential. The free massive potential is selected subsequently because it produces the linear equations and homogeneous quadratic Hamiltonians required for the explicit canonical transformation. An additive constant does not affect the reduced field equations, although it would contribute to the physical energy density if the background geometry were varied.

For the free massive field, the auxiliary equation contains the effective frequency squared $m^{2} - 3\dot{H}(t)$. Since $\chi$ is introduced as part of the enlarged variational system, the behavior of this auxiliary sector does not by itself represent a physical scalar-field instability. The same distinction applies to $H_{B,\mathrm{SF}}$, which is not the physical energy density of $\phi$. It is conserved when $H$ is constant but is not generically conserved on a nonconstant prescribed background; Eq.~\eqref{conservation_condition_Bateman_cosmological} characterizes the trajectories for which conservation nevertheless occurs.

The power-law background provides a concrete example of this trajectory-dependent conservation. For the background in Eq.~\eqref{power_law_background}, compatibility of the physical and auxiliary equations with the correlated trajectory \eqref{power_law_conservation_ansatz} selects the value in Eq.~\eqref{matter_power_law_special} for $m > 0$. Although $H(t)$ remains time dependent, Eqs.~\eqref{q_matter_power_law_definition}--\eqref{HB_matter_power_law_conserved} reduce the Bateman Hamiltonian to a conserved oscillator form. This conservation is a property of the correlated doubled trajectory on the prescribed background, rather than a consequence of a constant Hubble parameter.

Equations~\eqref{lagrangian_CK_phi_cosmological} and \eqref{lagrangian_CK_chi_cosmological} show how the CK factors generate the damped and anti-damped equations. For the canonical transformation, the auxiliary contribution in Eq.~\eqref{hamiltonian_CK_chi_cosmological} combines with the time-derivative terms in Eq.~\eqref{F2_time_derivative_cosmological} to yield the identity in Eq.~\eqref{Hamiltonian_equivalence_cosmological}. Under the linear point-transformation ansatz, the coefficient equations impose no additional condition on a sufficiently differentiable prescribed scale factor.

The canonical equivalence established here is an equivalence of reduced doubled Hamiltonian systems on a prescribed background. It does not establish equivalence of the physical one-field sectors and does not include the Friedmann constraint or the backreaction of the scalar field on $a(t)$. Consequently, the construction does not by itself determine a self-consistent expansion history, an equation-of-state evolution, or an observable dark-energy model.

Effective descriptions of dissipative inflation generate friction and noise through couplings to additional degrees of freedom~\cite{LopezNacirEtAl2012}. Harko~\cite{Harko2023} introduced an independent dissipation factor that modifies the scalar-field and Friedmann equations, whereas El-Nabulsi and Anukool~\cite{ElNabulsi2026} considered Bateman-type dual fields with nonlinear tachyonic and Born--Infeld dynamics. The construction considered here introduces neither an additional source of physical dissipation nor a nonlinear kinetic term. Instead, it gives an explicit, invertible time-dependent canonical map between the doubled CK and Bateman systems for a homogeneous massive scalar field with Hubble damping. For a general potential, the multiplier action determines the adjoint auxiliary dynamics, while the free massive specialization identifies the $\dot{H}(t)$ terms required for Hamiltonian equivalence on a prescribed FLRW background.
\section*{Acknowledgments}

This work was supported by Prince of Songkla University (Ref. No. SCI700130S-0), Thailand Science Research and Innovation (TSRI), and National Science, Research and Innovation Fund (NSRF).

\printbibliography

\end{document}